\documentclass[referee]{raa}            

\usepackage{graphicx,times}             
\usepackage{natbib}
\usepackage{amssymb,amsmath}
\usepackage{longtable}
\usepackage{threeparttable}
\usepackage{float}
\usepackage{CJK}
 \usepackage{threeparttable}
\usepackage{color}
\usepackage{epstopdf}
\usepackage{ragged2e} 
\usepackage{booktabs,makecell, multirow, tabularx}
\bibpunct{(}{)}{;}{a}{}{,}

\usepackage[a4paper=true,driverfallback=dvips,pagebackref=true]{hyperref}
\hypersetup{colorlinks = true, linkcolor = green, anchorcolor = red, citecolor = blue, filecolor = red, pagecolor = red, urlcolor = red}

\begin{document}
\begin{CJK*}{UTF8}{gbsn}

   \title{Distances to molecular clouds in the Galactic longitude $l=10\degr-20\degr$ from the MWISP $^{12}$CO 1-0 survey 
}

   \volnopage{}      
   \setcounter{page}{1}          

   \author{Juan Mei (梅娟)
          \inst{1,2}
          \and
          Zhiwei Chen (陈志维) \inst{3,4}\thanks{zwchen@pmo.ac.cn}
          \and
          Min Fang (房敏) \inst{1,2}\thanks{mfang@pmo.ac.cn}
          \and
          Miaomiao Zhang (张淼淼) \inst{1}
          \and   
          Shiyu Zhang (张世瑜) \inst{1}
          \and
          Zhibo Jiang (江治波) \inst{1}
   }

    \institute{Purple Mountain Observatory, Chinese Academy of Sciences, 10 Yuanhua Road, 210023 Nanjing, China\\
        \and
             University of Science and Technology of China, Chinese Academy of Sciences, Hefei 230026, China\\
        \and
            Center for Astronomy and Space Sciences, China Three Gorges University, 8 University Road, 443002 Yichang, China\\
        \and
            College of Science, China Three Gorges University, 8 University Road, 443002 Yichang, China
\vs\no
  {\small Received~~20xx month day; accepted~~20xx~~month day}}

\abstract{
We present distances to 56 molecular clouds within $10\degr \leq l \leq 20\degr$ and $|b| \leq 5.25\degr$ from the Milky Way Imaging Scroll Painting (MWISP) $^{12}$CO survey, 47 of which are first-time determinations. The molecular clouds were identified using the DBSCAN algorithm, and their distances were measured with the model-calibrated color-distance method using $J-K{_s}$ colors and the distances provided by 2MASS and \textit{Gaia} EDR3. The distances range from $\sim$275 pc to $\sim$2118 pc. We also derived the physical properties of molecular clouds and found a moderate correlation between the dust extinction and the $^{12}$CO integrated intensity.
}

   \authorrunning{Juan Mei et al.}            
  \titlerunning{Distances of molecular clouds}  

   \maketitle

%
%
\section{Introduction}           
Molecular clouds serve as the primary sites of star formation. Accurate distances are fundamental for deriving their intrinsic physical properties, such as mass and size. Therefore, determining molecular cloud distances is essential for understanding both star formation processes \citep{McKee+2007, Kennicutt+2012, HeyerDame+2015} and the Galactic structure \citep{Dame+2001, Xu+2018}. Although traditionally estimated from the Galactic rotation curve, such kinematic distances are suffer from large uncertainties and the near--far ambiguity problem, especially within the inner Galaxy.

The release of \textit{Gaia} has revolutionized distance estimation by providing precise stellar parallaxes, which enables molecular cloud distances to be determined from the extinction jump along the line of sight. Combined with \textit{Gaia} data, several studies have presented new three-dimensional (3D) extinction maps and estimated extinctions and distances of millions of stars \citep{Green+2019, ChenBQ+2019, Lallement+2019, Guo+2021, Sun+2021a, Wang+2025a}. These advances have facilitated successful distance determinations for molecular clouds across diverse Galactic environments. At high Galactic latitudes ($\lvert b \rvert \ge 10\degr$), \cite{Zucker+2019, Zucker+2020} determined distances to many molecular clouds and star--forming regions. Similarly, \cite{Yan+2019} derived distances to 11 molecular clouds, and \cite{Sun+2021b} determined the distances to 66 \citet[MBM]{Magnani+1985} molecular clouds based on the accurate color excess and stellar distance from ${\it Gaia}$. More recently, \cite{Sun+2024} identified 315 high-latitude clouds from Planck 857 GHz observations \citep{Planck+2020}, with distances derived from stellar parallax and extinction. 
Within the Galactic plane ($\lvert b \rvert \le 10\degr$), \cite{ChenBQ+2020a} used hierarchical clustering on 3D reddening maps from \cite{ChenBQ+2019} to identify and estimate distances of 567 dust clouds.  In a similar approach, \cite{Guo+2022} employed the same methodology to derive the distances of 250 dust clouds in the southern sky using extinction maps from \cite{Guo+2021}. Based on MWISP $^{12}$CO $J=1-0$ line emission, \cite{Yan+2021} applied DBSCAN to identify 2214 molecular clouds and measured distances for 234 of them using the background-eliminated extinction-parallax method. The fraction of molecular clouds with estimated distances is about 20$\%$ in the third Galactic quadrant but less than 10$\%$ in the first and second quadrants. Recently, \citet{Xie+2024} used the 3D dust map of \cite{Vergely+2022} to identify and estimate distances for 550 dust clouds within 3 kpc in the Galactic disk. \citet{Wang+2025b} constructed an all-sky catalog of 3345 dust clouds identified from their 3D dust-reddening map \citep{Wang+2025a} using dendrogram algorithm.

Despite these advances, accurate distance estimation for molecular clouds in the inner Galactic region remains challenging because it is hard to separate from the crowded field of clouds. Recently, \cite{Mei+2024} proposed a method for estimating the distances of molecular clouds based on the \textit{Gaia}, 2MASS and TRILEGAL galaxy model. This approach does not rely on an exact boundary of the cloud, making it particularly suitable for the crowded regions of the inner Galaxy. This method provide an opportunity to improve distance estimates for inner Galactic molecular clouds, which are crucial for revealing the detailed structure of the Galactic spiral arms and for understanding star formation under the extreme conditions of the inner Galactic region.

In this work, we measured distances to molecular clouds within the Galactic longitude range of $10\degr \leq l \leq 20\degr$ and the latitude range of $\lvert b \rvert \leq 5.25\degr$. The molecular clouds were identified in $^{12}$CO ($J=1-0$) data using the DBSCAN algorithm, and their distances were estimated using the model-calibrated color-distance (MCCD) method \citep{Mei+2024}. The MCCD approach searches for the color jump caused by a molecular cloud by comparing the reddening of stars projected on the cloud (on-cloud stars) with that of synthetic stellar populations generated by the TRILEGAL galaxy model. Since it is weakly dependent on the cloud boundary, the method is particularly suitable for measuring molecular clouds toward the inner Galactic region. Using this method, we determined distances for 56 molecular clouds, the most distant of which is ~2118 pc from the Sun.

The paper is structured as follows. In Section 2, we describe the data and method. Section 3 presents the distances of molecular clouds and discussions. We summarize in Section 4.

\section{Data and method}
\subsection{CO data}
The MWISP\footnote{\url{http://www.radioast.nsdc.cn/mwisp.php}} project is an ongoing CO survey by the Purple Mountain Observatory (PMO) 13.7 m millimeter telescope \citep[see more details of MWISP in][]{Su+2019}. The first data release of MWISP Phase I covers three CO isotopologue lines toward the northern Galactic plane within $9.75\degr \leq l \leq 229.25\degr$, $\lvert b \rvert \leq 5.25\degr$, which aims to study the molecular cloud properties, the star formation and the Galactic structures \citep{YangMWISP+2025}. 

In this paper, the $^{12}$CO image of the examined region ($10\degr \leq l \leq 20\degr$, $\lvert b \rvert \leq 5.25\degr$, and $-$16 $\leq$ V$_{\mathrm{LSR}}$ $\leq$ 32 km s$^{-1}$) is part of the MWISP CO survey. The MCCD method can measure distances up to 3 kpc due to the limitations of the \textit{Gaia} data, so we selected CO emission with V$_{\mathrm{LSR}}$ $\leq$ 32 km s$^{-1}$, containing the Sagittarius-Carina Arm and the Aquila Rift region. The rms noise of the $^{12}$CO emission is about 0.5 K at a velocity resolution of $0.16\,\mathrm{km^{-1}}$ and a grid spacing of $30\arcsec\times30\arcsec$.

\subsection{\textit{Gaia} EDR3 and 2MASS}
The \textit{Gaia} Early Data Release 3 (\textit{Gaia} EDR3; \citealt{Gaia3+2021}) consists of astrometry and photometry for 1.8 billion sources, featuring highly precise parallax measurements. We adopt stellar distances estimated by \cite{BailerJ+2021}, which transfers the \textit{Gaia} EDR3 parallaxes to distances using the Bayesian approach. The 2MASS \citep{Skrutskie+2MASS+2006} provided full coverage of the sky, which has three near-IR bands, $J$, $H$, and $K_s$, centered at 1.25, 1.65, and 2.16 $\mu m$, respectively. The systematic uncertainties of 2MASS photometric measurements are $<$0.03 mag. For each star, We can obtain its distance ($d$) from \textit{Gaia} EDR3 catalog and  $J-K{_s}$ color from 2MASS catalog \citep{Cutri+2MASScatalog+2003} with $J$ and $K_s$ bands. For each molecular cloud, we cross-matched \textit{Gaia} EDR3 catalog and 2MASS catalog within 1 arcsec for stars. The standard deviation of distance $\Delta d$ is calculated with the 16th and 84th percentiles of distance, which are given by \textit{Gaia} EDR3 catalog. The standard deviation of color is calculated by the error propagation.

\subsection{Molecular cloud identification}
\begin{figure*}
    \centering
	\includegraphics[width=0.95\linewidth]{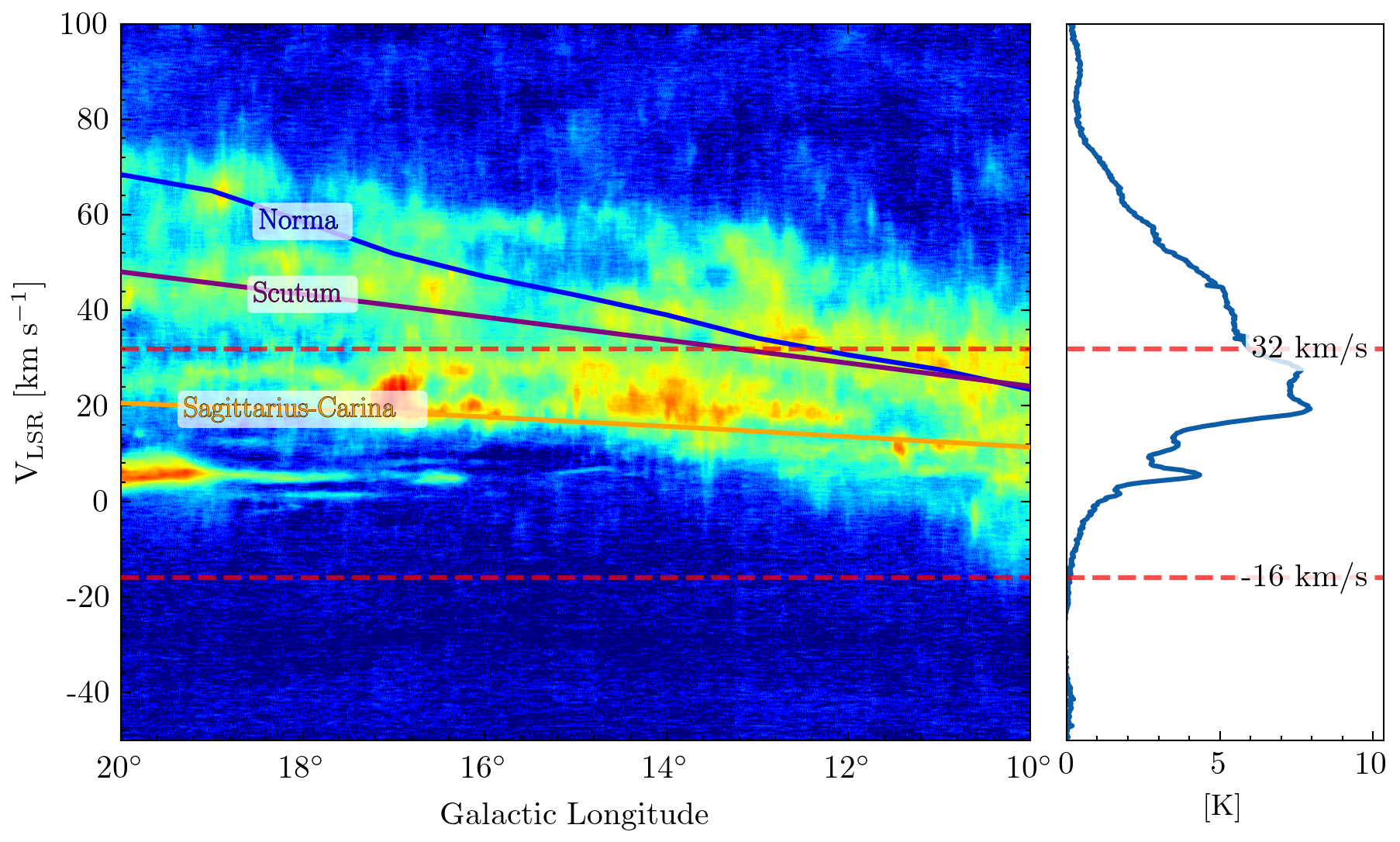}
    \caption{$l-v$ diagram (left) and average spectra (right) of $^{12}$CO emission. The colored solid lines represent different spiral arm models from \citet{Reid+2019}: Sagittarius--Carina (orange), Scutum (purple), and Norma (blue). The red dashed lines indicate the velocity range adopted in this work ($-16$ to $32$ km s$^{-1}$).}
    \label{lv_linev}
\end{figure*}

We apply the DBSCAN algorithm \citep{Ester+1996} to identify molecular clouds from the MWISP $^{12}$CO $J=1-0$ line emission within the Galactic longitude range of $l=10\degr$$-$$20\degr$. Due to the distance limitations of the MCCD method, which provides reliable distance estimates only within $\sim3$ kpc of the Sun \citep{Mei+2024}, primarily owing to the limited number of on-cloud stars with precise parallaxes, the insufficiency of reddened stars to produce a clear color jump, and the unresolved complexity of overlapping clouds along the line of sight.
Fig.~\ref{lv_linev} shows the $l-v$ diagram of $^{12}$CO emission overlaid with the spiral arm model of \citet{Reid+2019}. Within the examined region ($l=10\degr-20\degr$), the Aquila Rift region and Sagittarius-Carina Arm exhibit velocities predominantly in the range of approximately $-16$ to $32$ km s$^{-1}$, while other arms such as Scutum and Norma show significantly higher velocities corresponding to larger distances. Therefore, we focused on the molecular cloud samples identified within this velocity interval, which corresponds to the Sagittarius-Carina Arm and the Aquila Rift region.
The selected region lies in the inner Galaxy and contains two prominent massive star-forming regions, namely M16 \citep{Hill+2012} and M17 \citep{Chen+2021, Yin+2022}. 
A molecular cloud is a coherent structure composed of contiguous position--position--velocity (PPV) voxels above a given intensity threshold. 
Since DBSCAN identifies clouds based solely on their clustering in PPV space, it does not account for the kinematic distance ambiguity. It is possible that two distinct clouds at different distances with similar velocities and overlapping projections along the line of sight could be merged into a single identified structure. The distance estimates in this work are derived from the MCCD method, which provides a statistically most probable distance for the cloud as an integrated structure.
The DBSCAN algorithm has two parameters, MinPts and Eps. The Eps defines the neighborhood of voxels, while MinPts defines core points, which are voxels with at least MinPts neighboring voxels, including themselves. \citet{Yan+2020} tested the effectiveness and robustness of the DBSCAN algorithm against different sets of MinPts with the MWISP data and found that the parameter set of Eps=1 and MinPts = 4 is most suitable for the MWISP data. They applied the following selection criteria to remove noise from DBSCAN clusters: (1) the minimum voxel number is 16, (2) the minimum peak brightness temperature is 5$\sigma$, (3) the projection area contains a beam (a compact 2$\times$2 region), (4) the minimum channel number in the velocity axis is 3. The first two criteria are related to sensitivity, and the last two are related to resolution. In this work, we adopted the same four criteria as \citet{Yan+2020}.

\begin{figure*}
    \centering
    \includegraphics[width=0.8\linewidth]{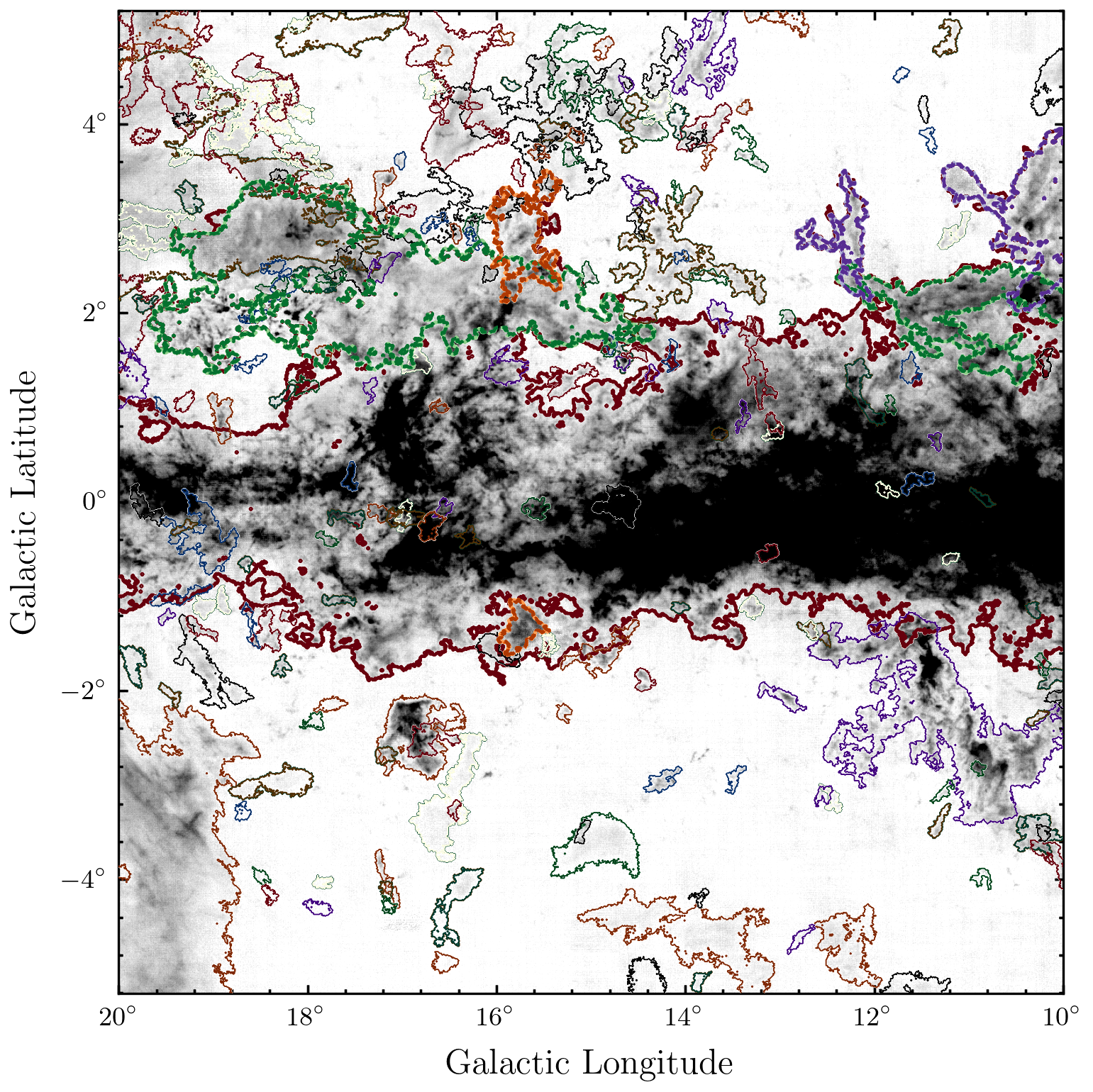}
    \caption{$l-b$ distribution of the identified molecular clouds. The background is the integrated intensity map of $^{12}$CO emission. The boundaries of the clouds are identified by the DBSCAN algorithm and outlined using different colors.
    }
    \label{cloud_samples}
\end{figure*}

The minimum cutoff on the PPV data cubes is 2$\sigma$ ($~$1 K). By setting the two DBSCAN parameters, Eps=1 and MinPts=4, we extracted 210 molecular clouds as shown in Fig. \ref{cloud_samples}. In Fig. \ref{cloud_samples}, we present the edges of 210 molecular clouds identified with different colored solid contours, with one large molecular cloud indicated by a thick solid red contour. The appearance of this very large molecular cloud around Galactic latitude 0$\degr$ is due to a limitation of the DBSCAN algorithm, which can pick up weak emissions but cannot properly separate large structured molecular clouds. We further processed this large molecular cloud by searching the entire cube by visually inspecting and identified six molecular clouds, as shown by the bold dashed contours in Fig. \ref{cloud_samples}. The boundaries are determined by selecting the velocity range corresponding to the target cloud before applying the DBSCAN algorithm. Since it is difficult to visually distinguish additional $^{12}$CO cloud structures, we applied the clustering to organize nested structures (ACORNS) algorithm to the $^{13}$CO data cube of the large molecular cloud. This algorithm, which is based on Gaussian decomposition and follows the method of \cite{Zhang+2024}, identified 3117 $^{13}$CO clouds.

In total, we identified 216 $^{12}$CO cloud samples and 3117 $^{13}$CO cloud samples in this region. The following section focuses on estimating the distances to these molecular clouds.

\subsection{Distance measurement}
For the identified $^{12}$CO and $^{13}$CO cloud samples, the distances to the $^{12}$CO clouds can be determined using the MCCD method. However, the $^{13}$CO clouds are located along the Galactic plane, where multiple overlapping structures along the line of sight currently make the MCCD method inapplicable. In such cases, their distances can be inferred indirectly by identifying associated star-forming regions.

The procedure for estimating distances to molecular clouds is identical to that described in \cite{Mei+2024}, with certain parameter settings adjusted according to the characteristics of the clouds. The method does not require an exact cloud boundary because it employs the TRILEGAL Galactic model to mimic the stellar population without cloud extinction. This simulation removes irregular variations of stellar color that are unrelated to the molecular cloud, thereby revealing the $J-K_s$ color jump point caused by the target cloud. 

Stars from the merged \textit{Gaia}-2MASS catalog are selected based on several criteria: (1) the stars are required to be located within the irregular boundary of the molecular cloud, (2) the relative uncertainty of the on-cloud star distance is lower than 10$\%$, (3) the intensity of the on-cloud stars falling in the integrated CO emission is greater than an intensity threshold. Additionally, we note that the large number of clusters within the molecular cloud may influence distance estimation. This is primarily because the member stars of a cluster have similar distances but a broad color distribution, which together complicate the identification of the color jump caused by the molecular cloud. When such a signature of cluster contamination is identified in the color--distance distribution, the relevant cluster member stars \citep{Hunt+2023} are removed. This procedure was applied, for example, to excise stars of the open cluster IC 4725 from the G013.9$-$04.5 cloud.

\begin{figure}[htbp]
    \centering
    \begin{minipage}{0.33\textwidth}
        \centering
        \includegraphics[width=\textwidth]{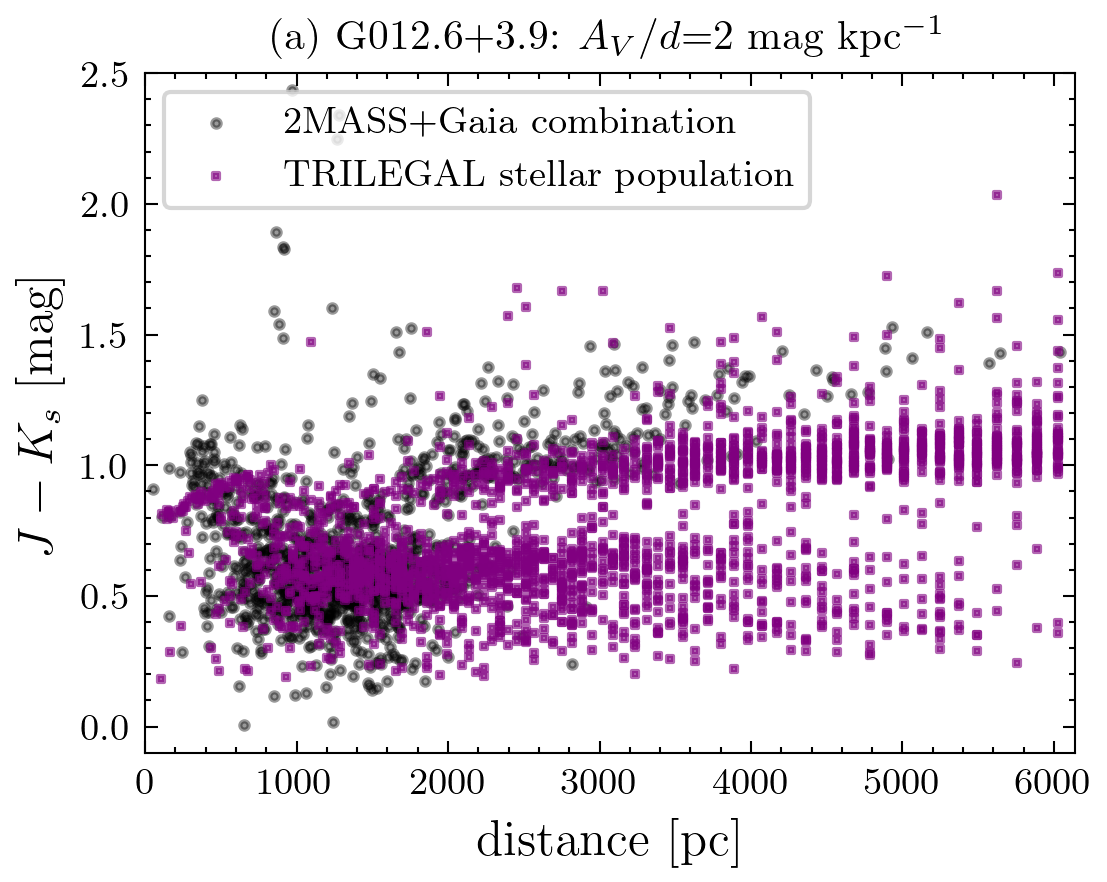}
    \end{minipage}
    \hfill
    \begin{minipage}{0.33\textwidth}
        \centering
        \includegraphics[width=\textwidth]{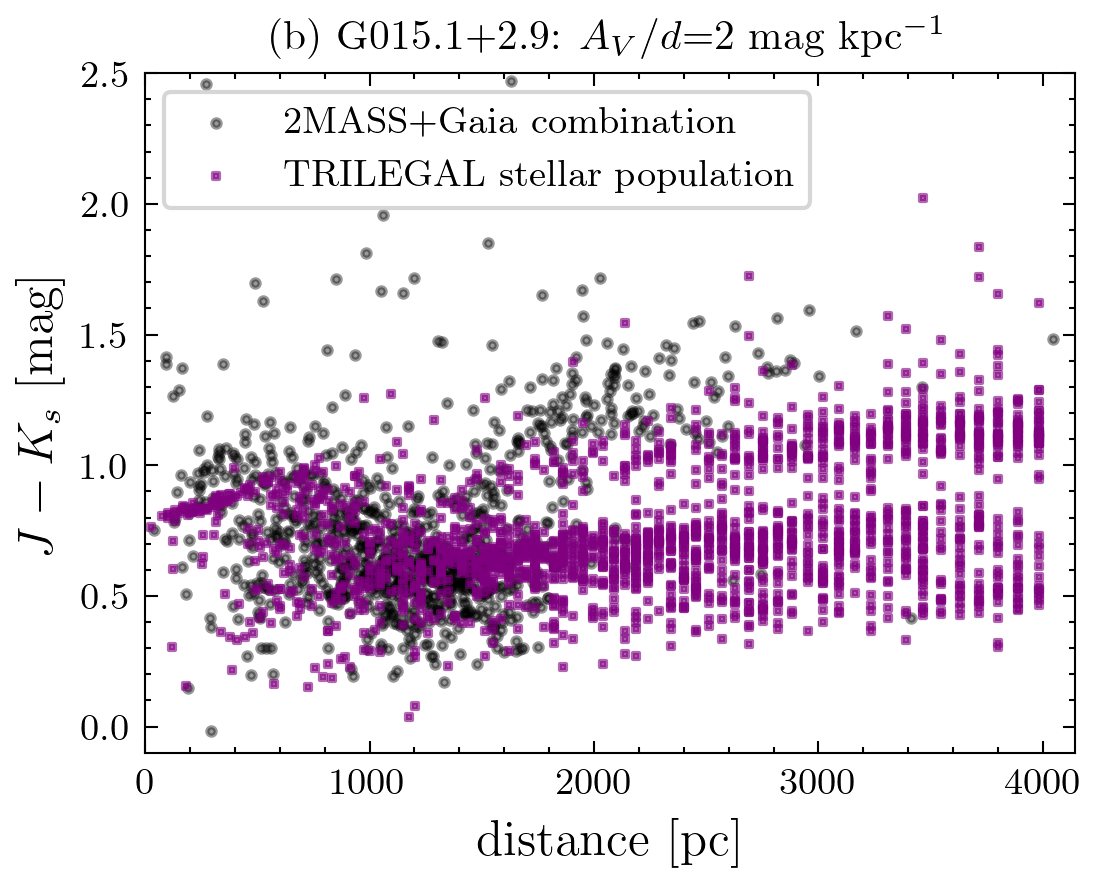}
    \end{minipage}    
    \hfill
    \begin{minipage}{0.33\textwidth}
        \centering
        \includegraphics[width=\textwidth]{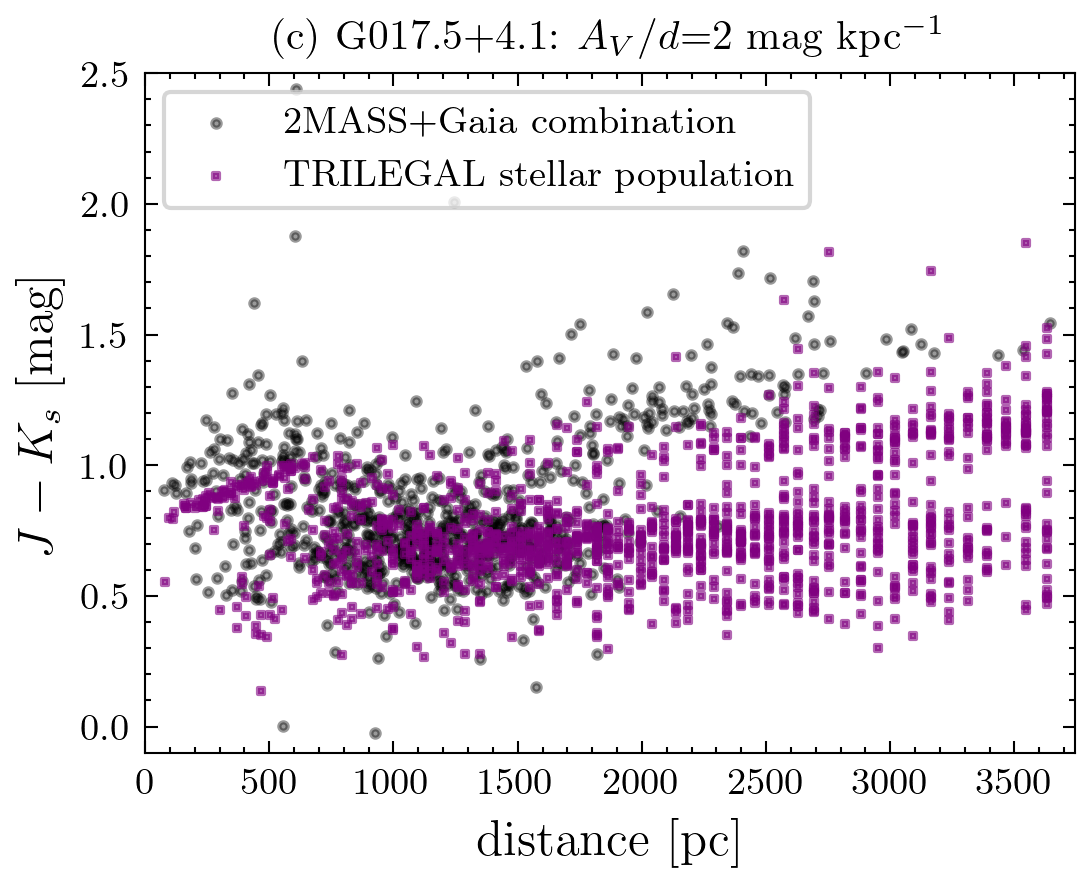}
    \end{minipage}
    
    \vskip 0.1cm

    \begin{minipage}{0.58\textwidth}
        \centering
        \includegraphics[width=\textwidth]{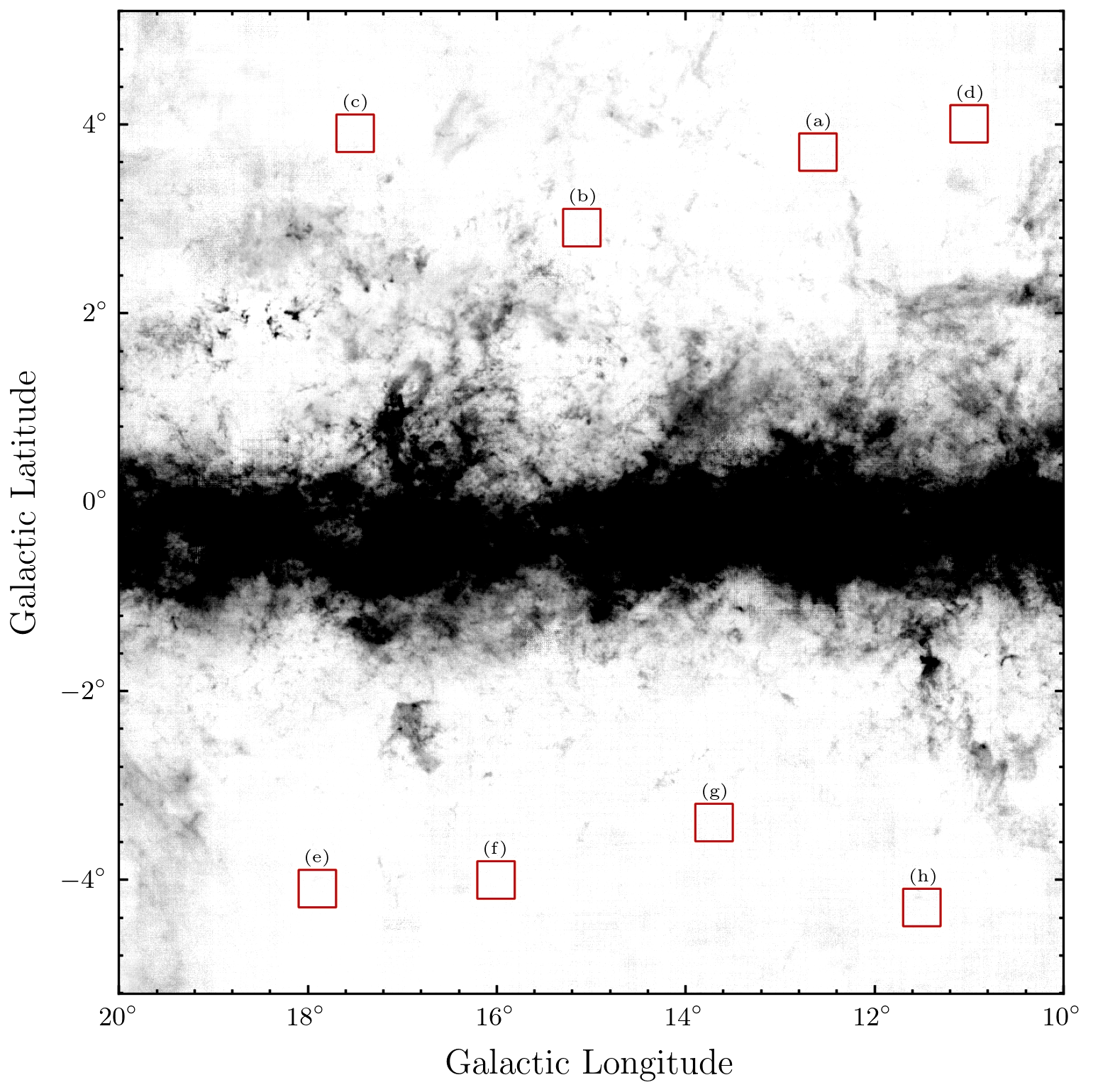}
    \end{minipage}
    \hfill
    \begin{minipage}{0.33\textwidth}
        \centering
        \includegraphics[width=\textwidth]{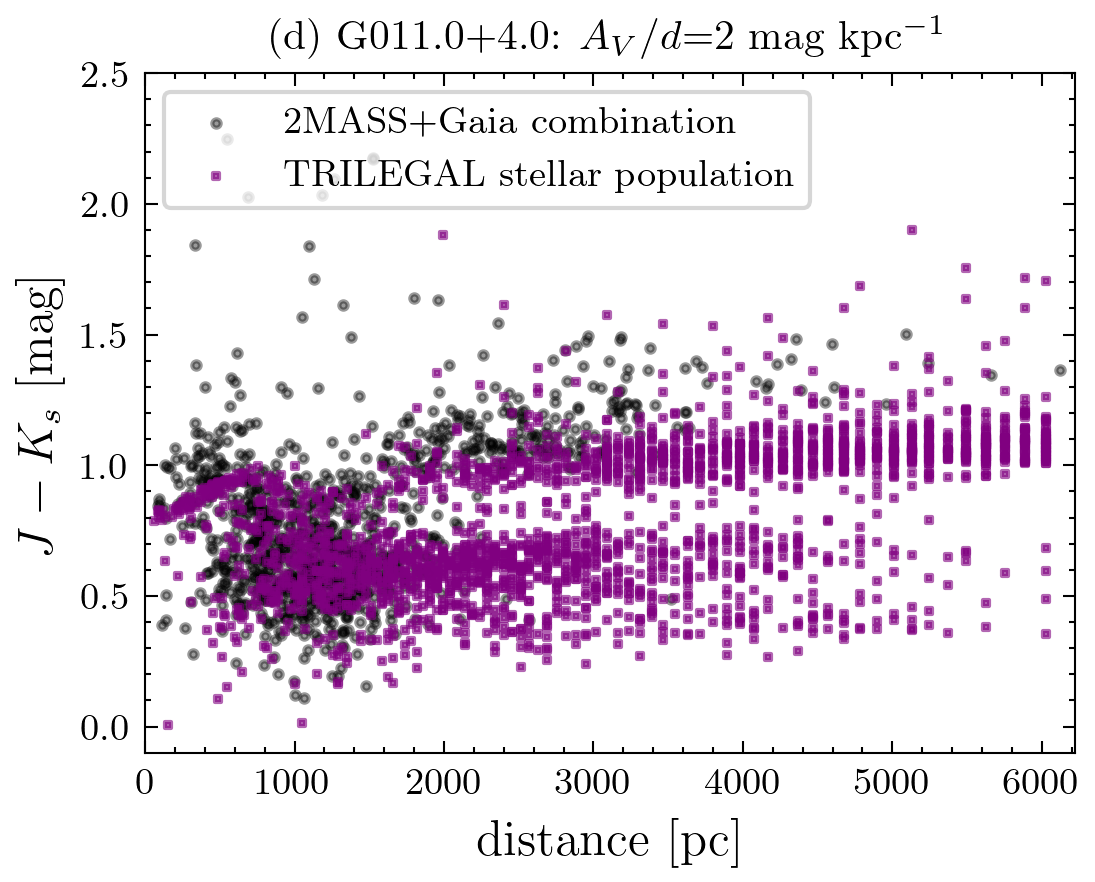}
        
        \vspace{0.2cm} 
        
        \includegraphics[width=\textwidth]{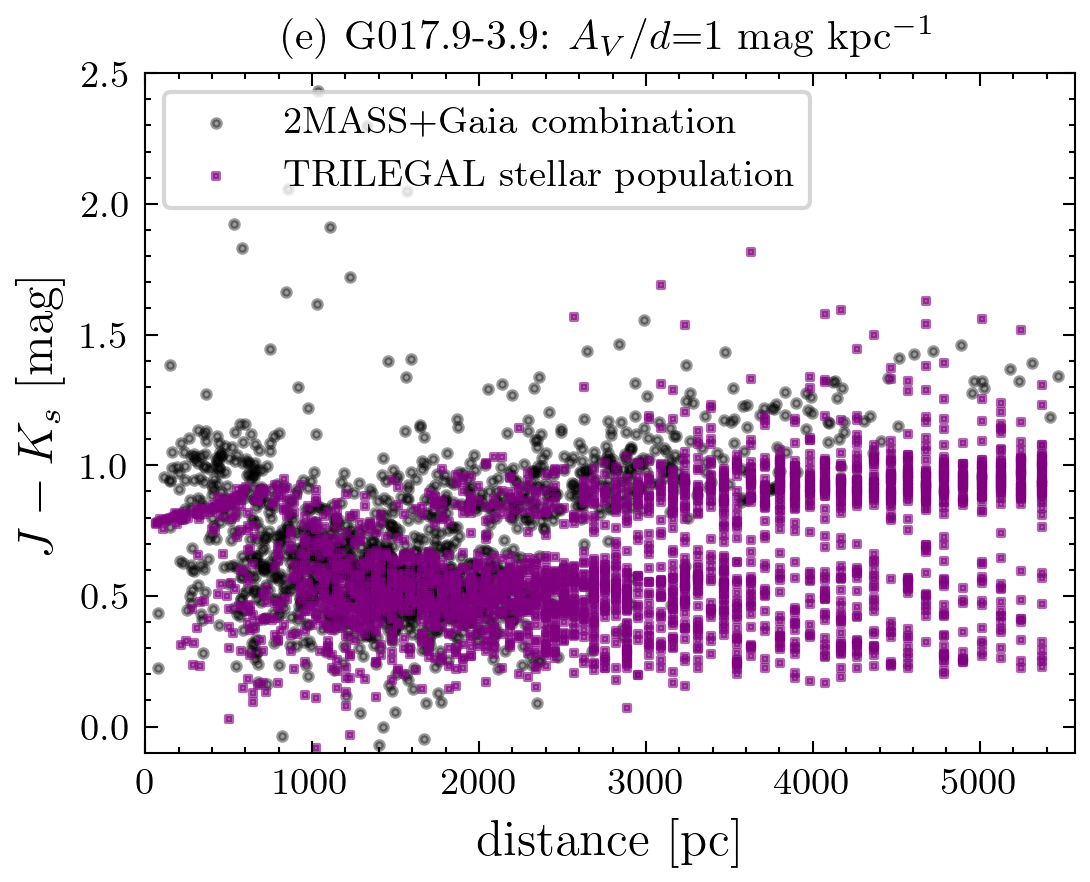}
    \end{minipage}
    
    \vskip 0.1cm
    
    \begin{minipage}{0.33\textwidth}
        \centering
        \includegraphics[width=\textwidth]{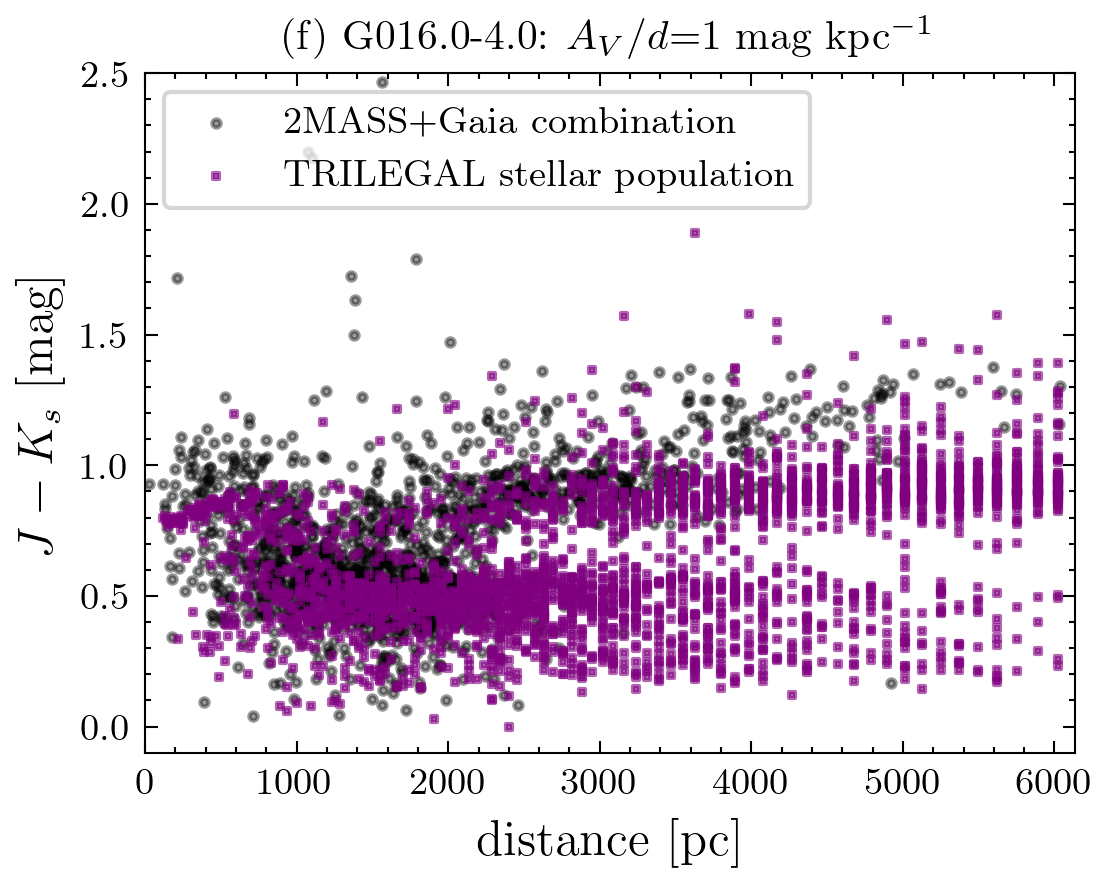}
    \end{minipage}
    \hfill
    \begin{minipage}{0.33\textwidth}
        \centering
        \includegraphics[width=\textwidth]{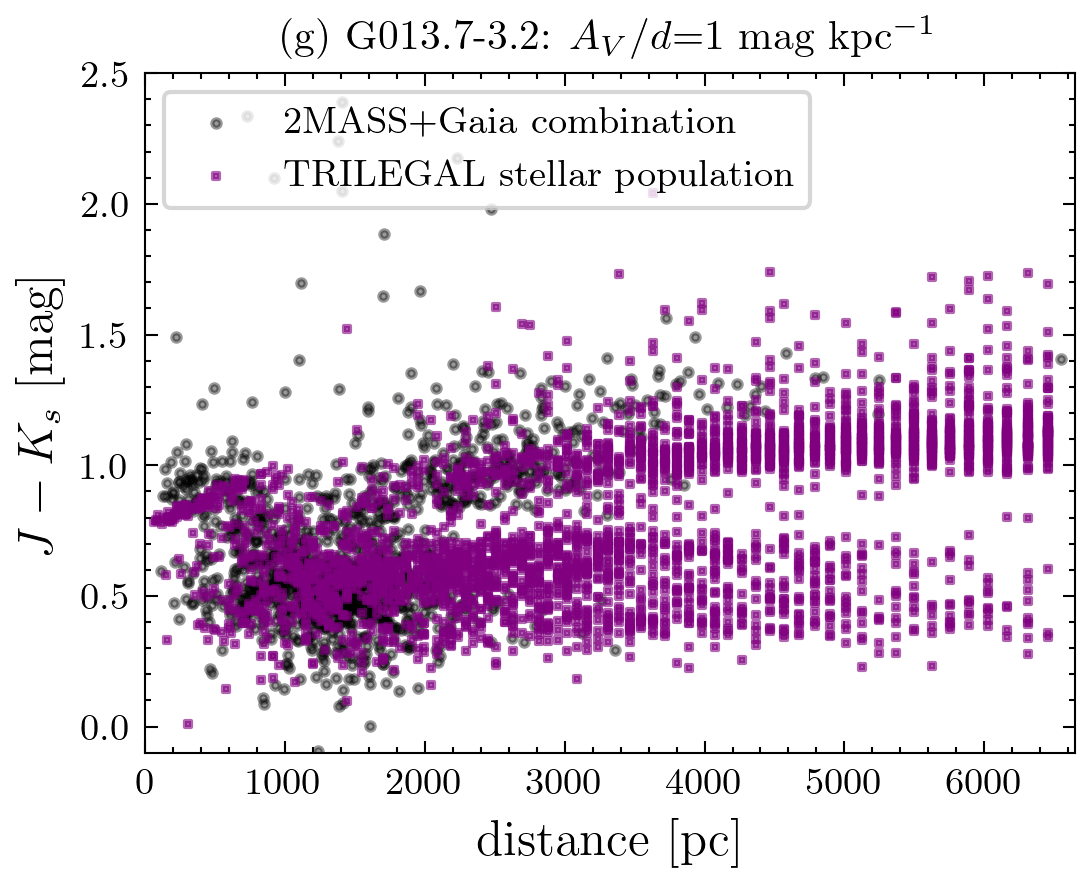}
    \end{minipage}
    \hfill
    \begin{minipage}{0.33\textwidth}
        \centering
        \includegraphics[width=\textwidth]{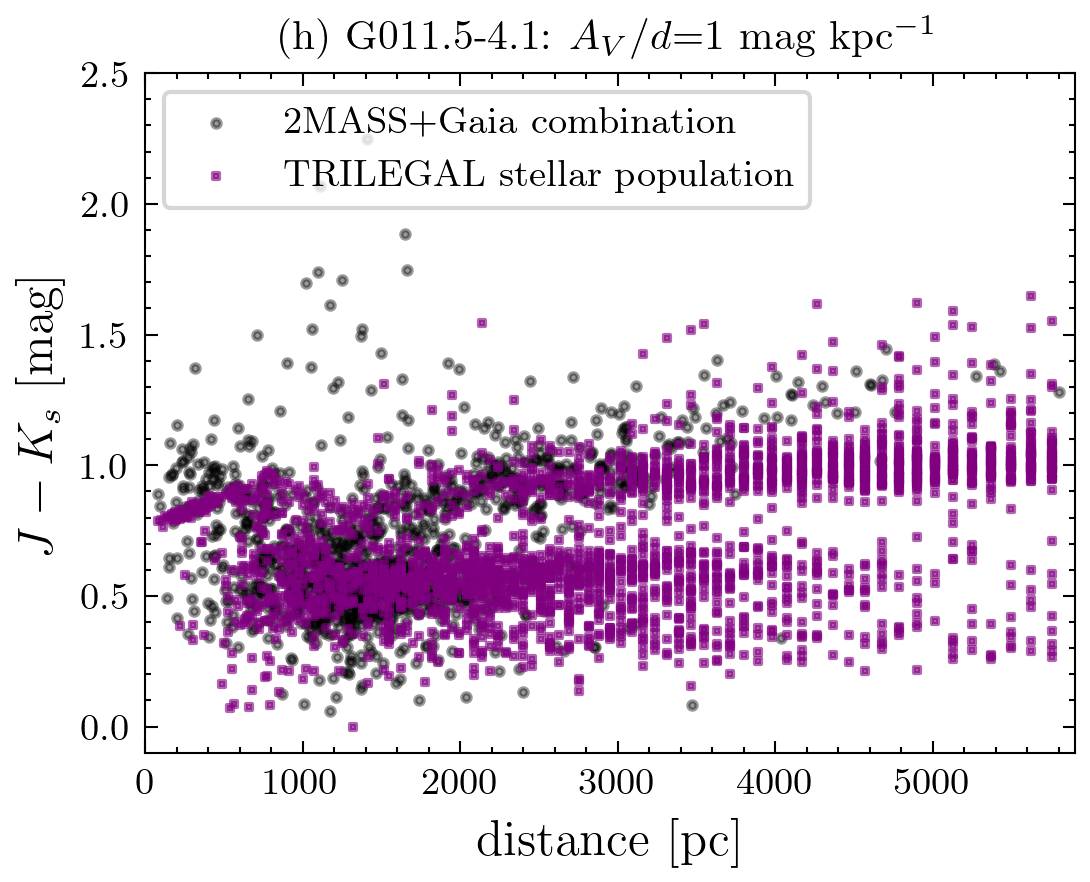}
    \end{minipage}

    \caption{The center-left panel displays the eight selected cloud-free areas (red boxes) overlaid on the integrated intensity map of $^{12}$CO emission. The remaining panels present the $J-K_s$ color vs. distance diagrams for the TRILEGAL stellar population (purple squares) and the 2MASS+Gaia combination (black points) within the selected cloud-free areas, respectively. The dust extinction trend in the upper four areas (a, b, c, d), located at Galactic latitude  $b > 0\degr$, follows $A_V/d = 2\,\mathrm{mag\,kpc}^{-1}$, whereas in the lower four areas (e, f, g, h), at $b < 0\degr$, the trend follows $A_V/d = 1\,\mathrm{mag\,kpc}^{-1}$.}
    \label{cloud_free_areas}
\end{figure}

The MCCD method \citep{Mei+2024} employed the TRILEGAL galaxy model to simulate the stellar population without cloud extinction along the sightline toward the cloud. For each cloud, we generated a collection of synthetic stellar populations within an area slightly larger than the cloud using the online input form of TRILEGAL 1.6 \footnote{\url{http://stev.oapd.inaf.it/cgi-bin/trilegal}}, adapting the available options to the molecular clouds in this region. We adjusted the limiting magnitude in the $K_s$ band for each cloud based on the $K_s$-band magnitude distribution of the on-cloud stars. To set the extinction parameter, we selected several areas free of $^{12}$CO $J=1-0$ line emission to mimic the diffuse ISM and adjusted their $K_s$-band limiting magnitudes accordingly. By adopting a dust extinction trend of $A_V/d = 2~\mathrm{mag~kpc}^{-1}$ for $b > 0\degr$ and $A_V/d = 1~\mathrm{mag~kpc}^{-1}$ for $b < 0\degr$, we found a good match in the color–distance diagram between the 2MASS+Gaia combination for cloud-free areas and the TRILEGAL stellar population, as detailed in Figure \ref{cloud_free_areas}. The other input parameters of TRILEGAL are follow those of \cite{Mei+2024}.

For each cloud, the $J-K_s$ color versus distance distribution of on-cloud stars is compared with that of the corresponding TRILEGAL synthetic population.
The intrinsic $J-K_s$ colors of main-sequence (MSs) and red giant stars (RGs) from TRILEGAL were considered separately and used as the baseline for subtracting the observed colors. 
The cloud distance is modeled as a switch point of two Gaussian distributions, from the foreground stellar color distribution (with a low mean value) to the background stellar color (with a high mean value). A Bayesian model is solved with Markov chain Monte Carlo (MCMC) sampling to determine the distance at which the $J-K_s$ color jump is most significant.

The distance determination procedure starts from the largest molecular clouds until the number of stars on the cloud is insufficient to find a significant color jump, and stops calculating when consecutively 50 molecular clouds have no distance detection. 
We retain only those molecular clouds that show unique and evident color jumps. If the distance is not successfully calculated, possible reasons include: the reddening of the stars due to the molecular cloud is not sufficient to cause a significant color jump, insufficient number of on-cloud stars, molecular clouds with severe foreground emission and molecular clouds at too great a distance ($>$3kpc).

\begin{figure*}
    \centering
	\includegraphics[width=0.9\linewidth]{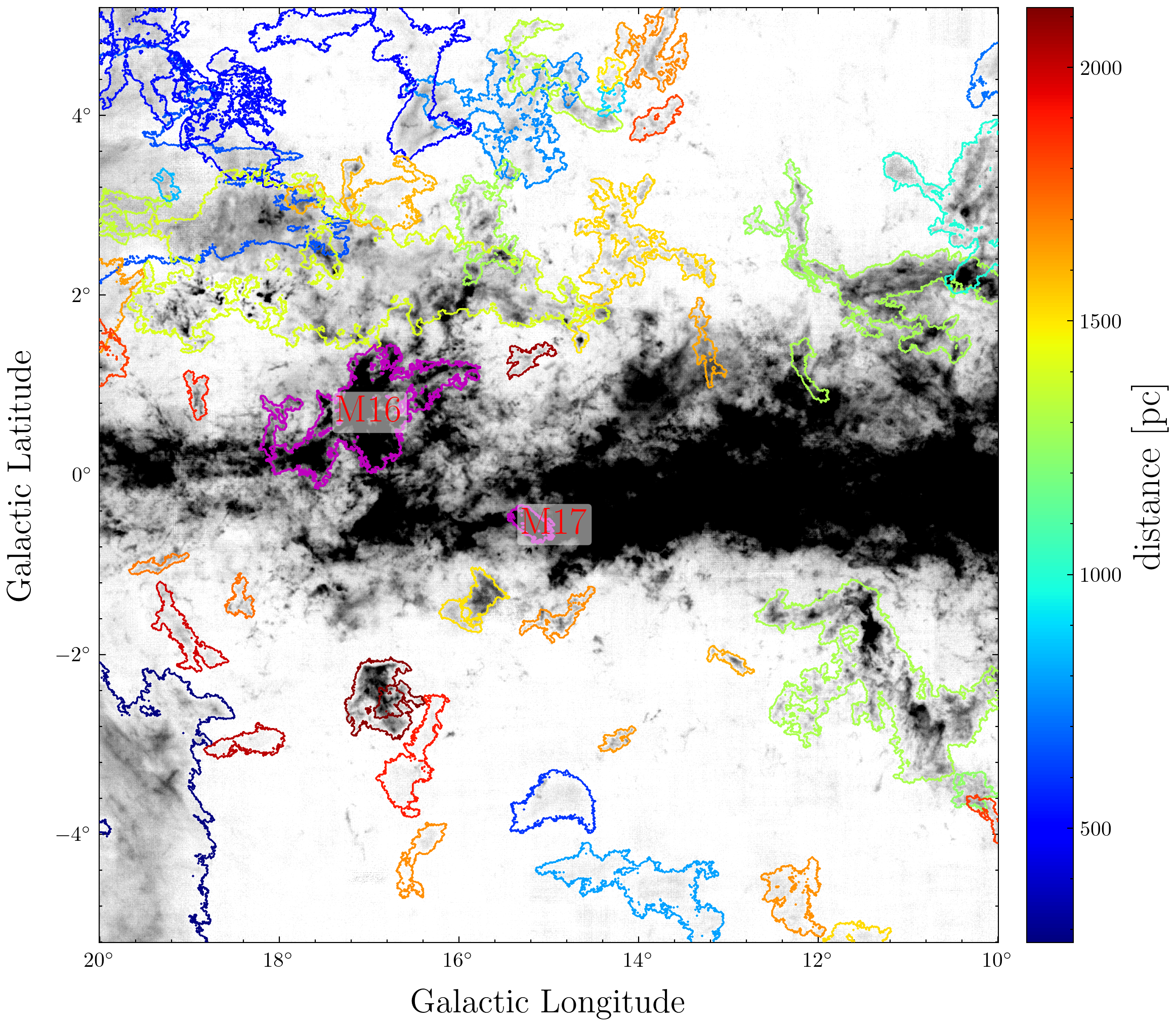}   
    \caption{The distances of 56 $^{12}$CO molecular clouds and two star-forming regions. The background is the integrated intensity map of $^{12}$CO emission. The contour colors correspond to the distances of their respective molecular clouds. The colored contours outline 56 $^{12}$CO molecular clouds, with colors corresponding to distances determined by the MCCD method. The magenta contours mark the boundaries of the M16 and M17 star-forming regions, which comprise multiple $^{13}$CO clouds.
    }
    \label{cloud_distances}
\end{figure*}

The systematic uncertainties from the MCCD method are approximately 10$\%$, and can be larger for distant molecular clouds. As discussed in \cite{Mei+2024}, stellar parallax errors can cause systematic shifts ($\sim$ 10$\%$) in the distance, but this effect is only significant for distant molecular clouds, whose background stars have large relative parallax errors. Other systematic errors arise from the systematic error of Gaia parallaxes, the choice of lower CO emission limits for on-cloud stars, and the unknown systematic errors of the Milky Way model. 
These systematic effects are not reflected in the statistical errors derived from the MCMC fitting procedure. Therefore, the total uncertainties of the distances estimated with the MCCD method should include both statistical and systematic uncertainties.

\section{Result and discussion}   
\subsection{Distances of the molecular clouds}
As shown in Fig. \ref{cloud_distances}, we examined 216 $^{12}$CO clouds and derived distances for 56 of them, while assigning literature-based distances to 129 $^{13}$CO clouds associated with the star-forming regions (M16 and M17). 

\begin{figure*}[!htbp]
    \centering
    \begin{subfigure}{\linewidth}
        \centering
        \includegraphics[width=1\linewidth]{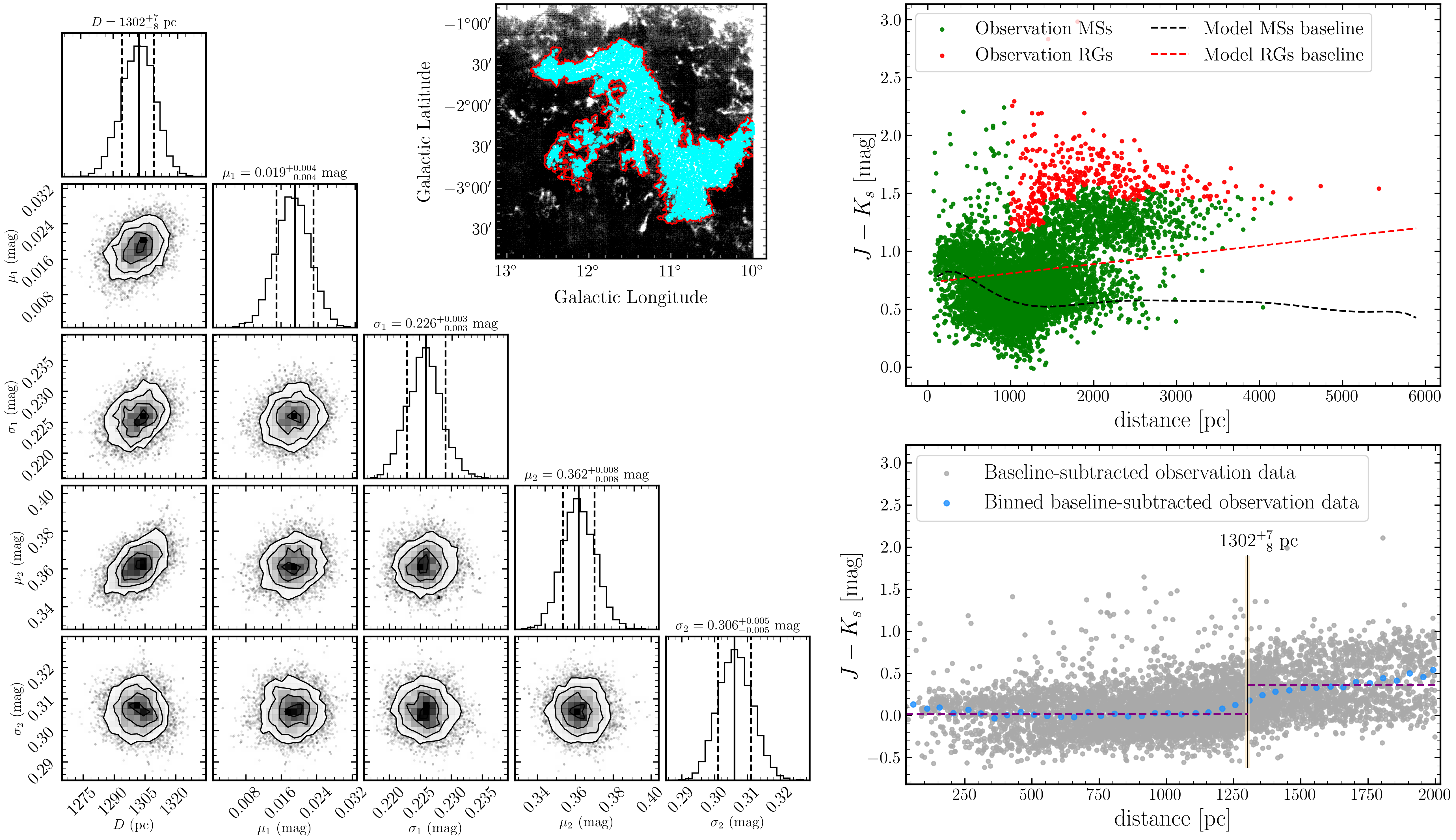}  
        \caption{}
        \label{cloud20550_distance}
    \end{subfigure}
    
    \vspace{0.3cm} 

    \begin{subfigure}{\linewidth}
        \centering
        \includegraphics[width=1\linewidth]{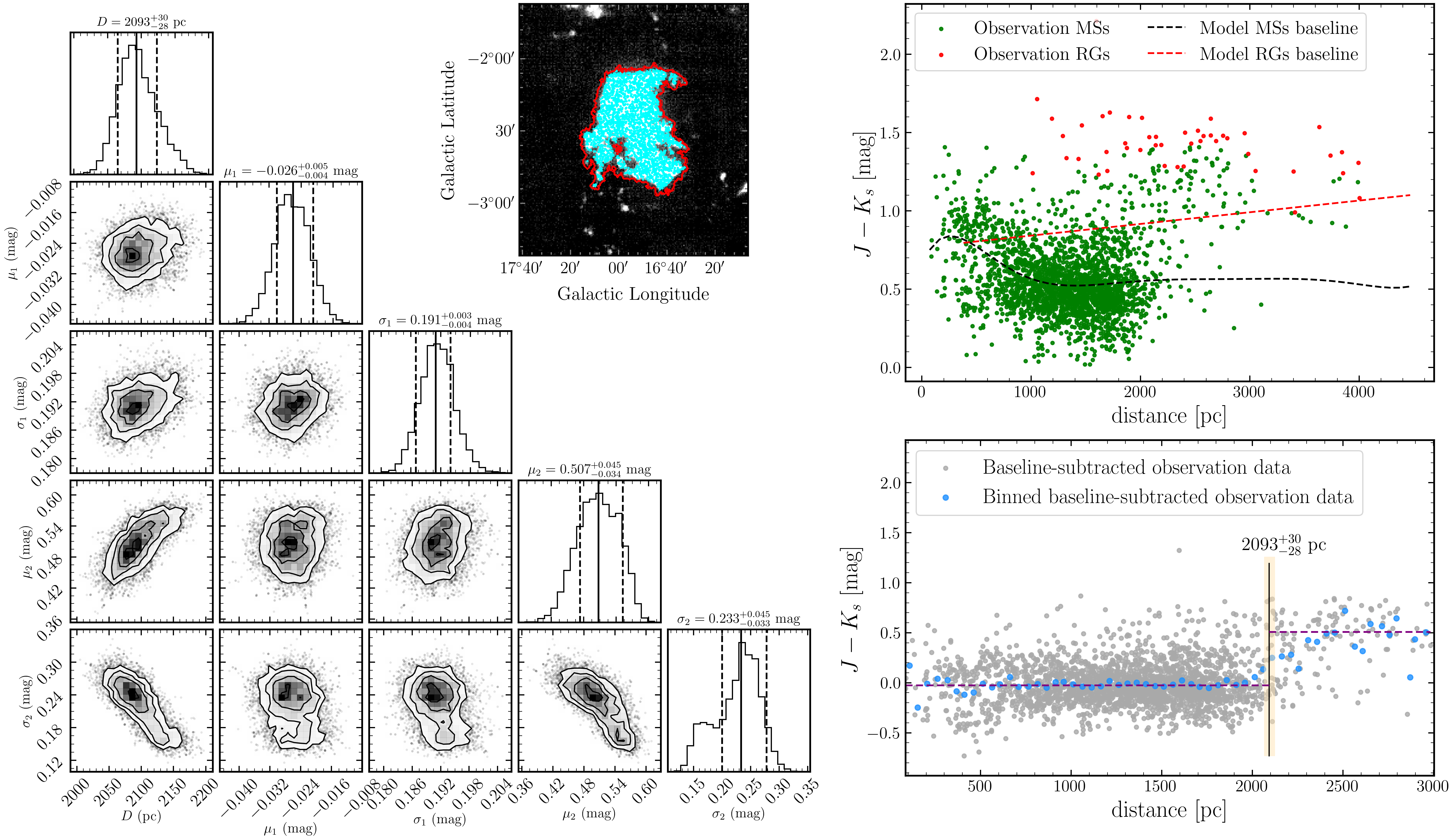}    
        \caption{}
        \label{cloud111715_distance}
    \end{subfigure}

    \caption{Distances of G011.2$-$02.2 (a) and G016.8$-$02.4 (b). The red contours refer to the molecular cloud footprint, and cyan points represent the starts projected on the cloud. The left and bottom right panels are corner maps and the estimated distance of the MCMC sampling. The black vertical lines indicate the estimated distance ($D$) and the shadow areas depict the 16th and 84th percentile values of the cloud distance. The broken horizontal purple dashed line is the color variation of on-cloud stars after subtracting the baseline. The red and green points in the top right panel present RGs and MSs distinguished in the observation data, respectively. The black and red dashed lines are the model-fitted RGs and MSs baseline, respectively.}
    \label{cloud20550a_111715b_distance}
\end{figure*}

The results of 56 molecular clouds are summarized in Table \ref{Catalog}. 
From left to right, we display the molecular cloud name (2), averaged l--b--V position (3, 4, 5), the angular area (6), the linear radius (7), the minimum CO emission of the on-cloud stars (8), the maximum distances to the on-cloud stars (9), the number of the on-cloud stars (10), the estimated distances of this work (11), molecular cloud mass (12), kinematic distances derived from the A5 model in \cite{Reid+2014} (13), derived from \cite{Reid+2019} (14) and notes (15). Among them, The molecular cloud mass were estimated by assuming a $^{12}$CO-to-H$_{2}$ mass conversion factor of X=2.0$\times$10$^{20}$ cm$^{-2}$ (K km s$^{-1})$ $^\mathrm{-1}$ \citep{Bauermeister+2013}. The typical statistical uncertainty of the distance is $\sim$ 5$\%$, and the systematic uncertainty is $\sim$ 10$\%$. The distance uncertainties reported in this work include both statistical and systematic uncertainties. The nearest distance to these molecular clouds is 275 pc, and the farthest distance is 2118 pc. The distance uncertainties might be larger for distant molecular clouds, such as G019.0$-$01.7 and G015.2$-$01.2. As examples, we display distances to two molecular clouds, G011.2$-$02.2 and G016.8$-$02.4 in Fig.~\ref{cloud20550a_111715b_distance}, and the other 54 molecular clouds are illustrated in the online supplementary material \footnote{\url{https://doi.org/10.57760/sciencedb.29146}}.

\begingroup
\renewcommand{\arraystretch}{0.9}
\setlength{\tabcolsep}{0.6 pt}
\scriptsize
\begin{longtable}{ccccccccccccccc}
    \captionsetup{justification=raggedright, singlelinecheck=false}
    \caption*{\normalsize \textbf{Table \thetable.} Distances to 56 molecular clouds.}\label{Catalog} \\
    \hline\noalign{\smallskip}
    \hline
    ID & Name & l & b & V$_{\rm LSR}$ & Area & r & CO$_{\rm cut}$ $^{\textcolor{blue}{1}}$ & D$_{\rm cut}$ & N $^{\textcolor{blue}{2}}$ & D $^{\textcolor{blue}{3}}$ & Mass $^{\textcolor{blue}{4}}$ & D$_{\rm Reid2014}$ $^{\textcolor{blue}{5}}$ & D$_{\rm Reid2019}$ $^{\textcolor{blue}{6}}$ & Notes \\
     & & ($\degr$) & ($\degr$) & (km s$^{-1}$) & (deg$^2$) & (pc) & (K km s$^{-1}$) & (pc) &  & (pc) & (10$^{3}$ M$_{\odot}$) & (kpc) & (kpc) & \\
    (1) & (2) & (3) & (4) & (5) & (6) & (7) & (8) & (9) &(10) &(11) &(12) &(13) &(14) & (15)  \\
    \hline
    \endfirsthead
    \multicolumn{15}{@{}l}{\normalsize \textbf{Table 1.} continued.} \\
    \hline\noalign{\smallskip}
    \hline
    ID & Name & l & b & V$_{\rm LSR}$ & Area & r & CO$_{\rm cut}$ $^{\textcolor{blue}{1}}$ & D$_{\rm cut}$ & N $^{\textcolor{blue}{2}}$ & D $^{\textcolor{blue}{3}}$ & Mass $^{\textcolor{blue}{4}}$ & D$_{\rm Reid2014}$ $^{\textcolor{blue}{5}}$ & D$_{\rm Reid2019}$ $^{\textcolor{blue}{6}}$ & Notes \\
     & & ($\degr$) & ($\degr$) & (km s$^{-1}$) & (deg$^2$) & (pc) & (K km s$^{-1}$) & (pc) &  & (pc) & (10$^{3}$ M$_{\odot}$) & (kpc) & (kpc) & \\
    (1) & (2) & (3) & (4) & (5) & (6) & (7) & (8) & (9) &(10) &(11) &(12) &(13) &(14)  & (15) \\
    \hline
    \endhead
    \hline\noalign{\smallskip}
    \endfoot
    \endlastfoot
    1 & G$010.0+04.4$	&	10.096	&	4.493	&	6.4	&	0.14	&	2.6	&	1.2	&	1500	&	457	&	710	$^{+82}_{-{82}}$	&	0.4	&	0.80 	$^{+1.01}_{-{0.22}}$	&	2.60 	$^{+2.40}_{-{2.40}}$ &	\\
    2 & G$010.1-03.7$	&	10.128	&	$-$3.734	&	28.8	&	0.06	&	4.5	&	1.5	&	2500	&	323	&	1832	$^{+187}_{-{187}}$	&	1.4	&	3.31 	$^{+0.50}_{-{0.60}}$	&	4.13 	$^{+1.91}_{-{1.91}}$ &	\\
    3 & G$010.2-03.5$	&	10.277	&	$-$3.523	&	$-$8.5	&	0.13	&	4.4	&	4	&	1500	&	458	&	1244	$^{+127}_{-{127}}$	&	2.7	&	17.52 	$^{+2.15}_{-{17.13}}$	&	1.55 	$^{+0.09}_{-{0.09}}$ &	\\
    4 & G$010.3+02.9$	&	10.319	&	2.968	&	4.3	&	0.86	&	4.2	&	5.6	&	1500	&	2188	&	989	$^{+100}_{-{100}}$	&	3.1	&	0.43 	$^{+1.08}_{-{0.24}}$	&	0.68 	$^{+0.60}_{-{0.60}}$ &	\\
    5 & G$010.9+02.0$	&	10.943	&	2.093	&	26.3	&	0.88	&	12.1	&	4	&	2000	&	1809	&	1294	$^{+130}_{-{130}}$	&	13.7	&	2.99 	$^{+0.52}_{-{0.63}}$	&	3.93 	$^{+1.81}_{-{1.81}}$ &	\\
    6 & G$011.2-02.2$	&	11.226	&	$-$2.206	&	11.6	&	2.2	&	19	&	6	&	2000	&	6908	&	1302	$^{+130}_{-{130}}$	&	88.5	&	1.44 	$^{+0.79}_{-{1.01}}$	&	1.25 	$^{+0.05}_{-{0.05}}$ & L291	\\
    7 & G$011.5-05.1$	&	11.563	&	$-$5.132	&	11.2	&	0.1	&	4.9	&	2	&	2000	&	419	&	1525	$^{+161}_{-{161}}$	&	1	&	1.35 	$^{+0.79}_{-{1.00}}$	&	2.78 	$^{+2.64}_{-{2.64}}$ &	\\
    8 & G$012.1+01.2$	&	12.151	&	1.216	&	6.3	&	0.09	&	3.8	&	2.8	&	2500	&	184	&	1303	$^{+150}_{-{150}}$	&	0.8	&	0.67 	$^{+0.90}_{-{0.20}}$	&	1.25 	$^{+0.05}_{-{0.05}}$ &	\\
    9 & G$012.2+02.8$	&	12.26	&	2.818	&	21.3	&	0.21	&	5.7	&	1.8	&	2000	&	693	&	1273	$^{+128}_{-{128}}$	&	2.3	&	2.37 	$^{+0.58}_{-{0.70}}$	&	3.01 	$^{+2.08}_{-{2.08}}$ &	\\
    10 & G$012.2-04.7$	&	12.283	&	$-$4.761	&	10.9	&	0.29	&	8.9	&	2.5	&	2000	&	639	&	1668	$^{+170}_{-{170}}$	&	3.3	&	1.26 	$^{+1.14}_{-{0.63}}$	&	1.25 	$^{+0.05}_{-{0.05}}$ &	\\
    11 & G$012.9-02.0$	&	12.948	&	$-$2.071	&	19.9	&	0.05	&	3.7	&	1.5	&	2000	&	225	&	1619	$^{+167}_{-{167}}$	&	1.7	&	2.16 	$^{+0.59}_{-{0.71}}$	&	1.22 	$^{+0.14}_{-{0.14}}$ &	\\
    12 & G$013.2+01.4$	&	13.261	&	1.468	&	12.9	&	0.11	&	5.2	&	1.5	&	2000	&	407	&	1623	$^{+168}_{-{168}}$	&	1.4	&	1.43 	$^{+0.70}_{-{0.86}}$	&	1.25 	$^{+0.06}_{-{0.06}}$ &	\\
    13 & G$013.7+04.7$	&	13.799	&	4.736	&	10.9	&	0.28	&	8.7	&	5.6	&	3000	&	601	&	1682	$^{+170}_{-{170}}$	&	7.7	&	1.18 	$^{+0.72}_{-{0.88}}$	&	0.29 	$^{+0.15}_{-{0.15}}$ &	\\
    14 & G$013.8+03.9$	&	13.812	&	3.938	&	21	&	0.13	&	6.4	&	3	&	3000	&	603	&	1821	$^{+184}_{-{184}}$	&	3.5	&	2.17 	$^{+0.56}_{-{0.66}}$	&	1.09 	$^{+0.13}_{-{0.13}}$ &	\\
    15 & G$013.9-04.5$	&	13.906	&	$-$4.594	&	11	&	0.73	&	6.7	&	3.5	&	1500	&	701	&	796	$^{+85}_{-{85}}$	&	1.7	&	1.16 	$^{+0.72}_{-{0.88}}$	&	0.29 	$^{+0.15}_{-{0.15}}$ &	\\
    16 & G$014.0+02.5$	&	14.094	&	2.576	&	15.9	&	0.68	&	12.4	&	4.5	&	2000	&	1897	&	1529	$^{+155}_{-{155}}$	&	14.4	&	1.68 	$^{+0.63}_{-{0.75}}$	&	1.15 	$^{+0.14}_{-{0.14}}$ &	\\
    17 & G$014.1+04.8$	&	14.158	&	4.839	&	27.1	&	0.05	&	3.6	&	2.5	&	2500	&	186	&	1647	$^{+174}_{-{174}}$	&	0.8	&	2.62 	$^{+0.48}_{-{0.56}}$	&	3.50 	$^{+1.67}_{-{1.67}}$ &	\\
    18 & G$014.2-02.9$	&	14.237	&	$-$2.957	&	25.4	&	0.05	&	3.6	&	2.1	&	2500	&	391	&	1655	$^{+173}_{-{173}}$	&	1.5	&	2.08 	$^{+0.56}_{-{0.66}}$	&	3.29 	$^{+1.83}_{-{1.83}}$ &	\\
    19 & G$014.2+04.1$	&	14.289	&	4.186	&	20.5	&	0.06	&	2.2	&	1.8	&	1500	&	243	&	889	$^{+94}_{-{94}}$	&	0.4	&	2.48 	$^{+0.50}_{-{0.59}}$	&	2.37 	$^{+2.44}_{-{2.44}}$ &	\\
    20 & G$014.3+04.4$	&	14.312	&	4.45	&	18	&	0.06	&	3.6	&	3.5	&	2000	&	149	&	1493	$^{+156}_{-{156}}$	&	0.8	&	1.86 	$^{+0.59}_{-{0.70}}$	&	2.45 	$^{+2.51}_{-{2.51}}$ &	\\
    21 & G$014.6+01.5$	&	14.66	&	1.547	&	13.7	&	0.04	&	3.1	&	3.5	&	2000	&	161	&	1505	$^{+163}_{-{163}}$	&	1.4	&	1.42 	$^{+0.65}_{-{0.78}}$	&	1.23 	$^{+0.10}_{-{0.10}}$ &	\\
    22 & G$014.8+04.2$	&	14.845	&	4.277	&	28.6	&	0.42	&	8.5	&	4.2	&	1700	&	816	&	1345	$^{+139}_{-{139}}$	&	6.2	&	2.66 	$^{+0.46}_{-{0.54}}$	&	3.51 	$^{+1.66}_{-{1.66}}$ &	\\
    23 & G$014.8-01.6$	&	14.876	&	$-$1.614	&	16	&	0.14	&	6.1	&	3.5	&	2500	&	532	&	1673	$^{+169}_{-{169}}$	&	4.3	&	1.62 	$^{+0.61}_{-{0.73}}$	&	1.26 	$^{+0.15}_{-{0.15}}$ &	\\
    24 & G$014.9-03.6$	&	14.906	&	$-$3.614	&	6.5	&	0.36	&	3.6	&	4.4	&	1200	&	233	&	599	$^{+77}_{-{77}}$	&	0.6	&	0.59 	$^{+0.78}_{-{0.18}}$	&	0.29 	$^{+0.15}_{-{0.15}}$ &	\\
    25 & G$015.2+01.2$	&	15.242	&	1.285	&	29.4	&	0.09	&	6.2	&	3	&	3000	&	508	&	2067	$^{+216}_{-{216}}$	&	3.5	&	2.68 	$^{+0.45}_{-{0.52}}$	&	2.63 	$^{+1.16}_{-{1.16}}$ &	\\
    26 & G$015.4+04.0$	&	15.495	&	4.079	&	13.4	&	1.03	&	7.6	&	4.9	&	1500	&	1136	&	760	$^{+78}_{-{78}}$	&	3.9	&	1.34 	$^{+0.63}_{-{0.75}}$	&	0.31 	$^{+0.16}_{-{0.16}}$ &	\\
    27 & G$015.7+02.7$	&	15.72	&	2.777	&	20.3	&	0.43	&	4.5	&	3.5	&	1700	&	993	&	1292	$^{+133}_{-{133}}$	&	1.8	&	1.94 	$^{+0.54}_{-{0.63}}$	&	1.12 	$^{+0.13}_{-{0.13}}$ &	\\
    28 & G$015.7-01.3$	&	15.733	&	$-$1.318	&	16.2	&	0.11	&	5	&	1.8	&	2000	&	563	&	1506	$^{+153}_{-{153}}$	&	2.4	&	1.58 	$^{+0.59}_{-{0.70}}$	&	1.31 	$^{+0.21}_{-{0.21}}$ &	\\
    29 & G$016.0-01.5$	&	16.024	&	$-$1.557	&	30.8	&	0.08	&	4.1	&	1.5	&	2000	&	372	&	1519	$^{+155}_{-{155}}$	&	1	&	2.69 	$^{+0.44}_{-{0.50}}$	&	2.92 	$^{+0.53}_{-{0.53}}$ &	\\
    30 & G$016.3-04.1$	&	16.383	&	$-$4.135	&	4.9	&	0.14	&	6.1	&	1.2	&	2500	&	721	&	1677	$^{+174}_{-{174}}$	&	1.5	&	0.37 	$^{+0.76}_{-{0.17}}$	&	0.28 	$^{+0.15}_{-{0.15}}$ &	\\
    31 & G$016.4+04.2$	&	16.484	&	4.202	&	4.9	&	0.77	&	4.1	&	3	&	1000	&	640	&	478	$^{+49}_{-{49}}$	&	1.3	&	0.39 	$^{+0.75}_{-{0.17}}$	&	0.27 	$^{+0.15}_{-{0.15}}$ &	\\
    32 & G$016.5-03.1$	&	16.519	&	$-$3.175	&	4.6	&	0.32	&	10.6	&	2.8	&	2500	&	783	&	1895	$^{+192}_{-{192}}$	&	3.6	&	0.33 	$^{+0.76}_{-{0.17}}$	&	0.27 	$^{+0.15}_{-{0.15}}$ &	\\
    33 & G$016.7-02.4$	&	16.721	&	$-$2.490	&	28.5	&	0.11	&	6.8	&	1.2	&	3000	&	821	&	2118	$^{+215}_{-{215}}$	&	5.8	&	2.47 	$^{+0.45}_{-{0.52}}$	&	3.18 	$^{+1.74}_{-{1.74}}$ &  L379	\\
    34 & G$016.8+03.0$	&	16.831	&	3.049	&	5.4	&	0.38	&	9.7	&	2.4	&	2000	&	560	&	1594	$^{+162}_{-{162}}$	&	3.5	&	0.44 	$^{+0.73}_{-{0.17}}$	&	0.27 	$^{+0.15}_{-{0.15}}$ &	\\
    35 & G$016.8-02.4$	&	16.897	&	$-$2.428	&	17.9	&	0.44	&	13.6	&	5.6	&	3000	&	2469	&	2093	$^{+211}_{-{211}}$	&	55.1	&	1.65 	$^{+0.55}_{-{0.64}}$	&	1.14 	$^{+0.13}_{-{0.13}}$ &	\\
    36 & G$017.1+03.2$	&	17.193	&	3.298	&	20.4	&	0.07	&	4.2	&	4.9	&	2000	&	150	&	1592	$^{+163}_{-{163}}$	&	2.1	&	1.85 	$^{+0.52}_{-{0.60}}$	&	1.13 	$^{+0.13}_{-{0.13}}$ &	\\
    37 & G$017.4+02.4$	&	17.479	&	2.416	&	10.8	&	0.13	&	5.1	&	2.1	&	2000	&	446	&	1412	$^{+143}_{-{143}}$	&	4.5	&	0.98 	$^{+0.63}_{-{0.74}}$	&	1.23 	$^{+0.10}_{-{0.10}}$ &	\\
    38 & G$017.5+04.9$	&	17.564	&	4.987	&	4.6	&	0.32	&	2.9	&	2.4	&	1000	&	149	&	522	$^{+69}_{-{69}}$	&	0.2	&	0.34 	$^{+0.72}_{-{0.17}}$	&	0.27 	$^{+0.15}_{-{0.15}}$ &	\\
    39 & G$017.5+02.3$	&	17.593	&	2.324	&	27.8	&	4.7	&	30.1	&	4.2	&	2000	&	13517	&	1409	$^{+141}_{-{141}}$	&	123	&	2.36 	$^{+0.45}_{-{0.81}}$	&	2.00 	$^{+0.08}_{-{0.08}}$ & NGC6604	\\
    40 & G$017.7+03.0$	&	17.711	&	3.072	&	10.3	&	0.1	&	4.9	&	1.8	&	2000	&	282	&	1602	$^{+167}_{-{167}}$	&	2.8	&	0.93 	$^{+0.63}_{-{0.74}}$	&	1.19 	$^{+0.13}_{-{0.13}}$ &	\\
    41 & G$018.3-02.9$	&	18.329	&	$-$2.980	&	8.1	&	0.16	&	7.9	&	0.8	&	2500	&	1156	&	2018	$^{+204}_{-{204}}$	&	2.1	&	0.68 	$^{+0.65}_{-{0.16}}$	&	0.26 	$^{+0.13}_{-{0.13}}$ &	\\
    42 & G$018.4-01.4$	&	18.43	&	$-$1.423	&	18.7	&	0.07	&	4.6	&	3.5	&	2500	&	294	&	1734	$^{+181}_{-{181}}$	&	1.8	&	1.63 	$^{+0.52}_{-{0.60}}$	&	1.50 	$^{+0.04}_{-{0.04}}$ &	\\
    43 & G$018.4+04.4$	&	18.479	&	4.473	&	11.2	&	0.17	&	2	&	1	&	1500	&	415	&	485	$^{+52}_{-{52}}$	&	0.2	&	0.99 	$^{+0.60}_{-{0.70}}$	&	0.30 	$^{+0.15}_{-{0.15}}$ &	\\
    44 & G$018.6+03.7$	&	18.614	&	3.706	&	12.4	&	0.65	&	4	&	2.5	&	1000	&	611	&	503	$^{+51}_{-{51}}$	&	1	&	1.09 	$^{+0.59}_{-{0.68}}$	&	0.30 	$^{+0.15}_{-{0.15}}$ &	\\
    45 & G$018.6+04.1$	&	18.657	&	4.156	&	1.6	&	0.48	&	3.6	&	1.2	&	1400	&	924	&	529	$^{+60}_{-{60}}$	&	0.3	&	15.81 	$^{+0.18}_{-{15.07}}$	&	0.26 	$^{+0.14}_{-{0.14}}$ &	\\
    46 & G$018.9+00.8$	&	18.911	&	0.888	&	18.6	&	0.08	&	5.3	&	4	&	3000	&	356	&	1884	$^{+192}_{-{192}}$	&	4.5	&	1.60 	$^{+0.52}_{-{0.59}}$	&	1.49 	$^{+0.08}_{-{0.08}}$ &	\\
    47 & G$019.0-01.7$	&	19.059	&	$-$1.760	&	0.3	&	0.22	&	9.2	&	3	&	2700	&	1231	&	1988	$^{+204}_{-{204}}$	&	6.3	&	0.43 	$^{+15.68}_{-{0.15}}$	&	0.24 	$^{+0.14}_{-{0.14}}$ &	\\
    48 & G$019.2+03.2$	&	19.25	&	3.209	&	21.4	&	0.05	&	1.9	&	1.8	&	1500	&	171	&	840	$^{+104}_{-{104}}$	&	0.3	&	1.80 	$^{+0.49}_{-{0.56}}$	&	1.46 	$^{+3.23}_{-{3.23}}$ &	\\
    49 & G$019.3+03.3$	&	19.333	&	3.372	&	5	&	3.64	&	12.3	&	3	&	1000	&	3797	&	655	$^{+66}_{-{66}}$	&	12.1	&	0.36 	$^{+0.67}_{-{0.16}}$	&	0.27 	$^{+0.15}_{-{0.15}}$ &	\\
    50 & G$019.3-01.0$	&	19.372	&	$-$1.009	&	18.7	&	0.06	&	4.1	&	2.5	&	3000	&	273	&	1722	$^{+193}_{-{193}}$	&	1.2	&	1.58 	$^{+0.51}_{-{0.58}}$	&	1.50 	$^{+0.04}_{-{0.04}}$ &	\\
    51 & G$019.4-03.7$	&	19.493	&	$-$3.725	&	6.8	&	3.26	&	4.9	&	4.5	&	500	&	1259	&	275	$^{+28}_{-{28}}$	&	4.5	&	0.53 	$^{+0.64}_{-{0.16}}$	&	0.25 	$^{+0.13}_{-{0.13}}$ &	\\
    52 & G$019.5+04.7$	&	19.551	&	4.767	&	14.4	&	0.66	&	4.3	&	3.5	&	1000	&	577	&	534	$^{+54}_{-{54}}$	&	1.5	&	1.23 	$^{+0.55}_{-{0.63}}$	&	0.31 	$^{+0.16}_{-{0.16}}$ &	\\
    53 & G$019.6+02.8$	&	19.694	&	2.898	&	25.3	&	0.19	&	5.8	&	1.8	&	1600	&	499	&	1362	$^{+141}_{-{141}}$	&	1.6	&	2.05 	$^{+0.45}_{-{0.51}}$	&	2.00 	$^{+0.08}_{-{0.08}}$ &	\\
    54 & G$019.7+05.0$	&	19.755	&	5.018	&	4.5	&	0.27	&	2.7	&	2.5	&	1000	&	297	&	534	$^{+56}_{-{56}}$	&	0.5	&	0.31 	$^{+0.67}_{-{0.16}}$	&	0.27 	$^{+0.15}_{-{0.15}}$ &	\\
    55 & G$019.8+01.8$	&	19.891	&	1.879	&	2.8	&	0.25	&	8.1	&	2	&	2000	&	1005	&	1648	$^{+169}_{-{169}}$	&	3.7	&	0.12 	$^{+15.60}_{-{0.69}}$	&	0.24 	$^{+0.15}_{-{0.15}}$ &	\\
    56 & G$019.9+01.2$	&	19.914	&	1.247	&	30.7	&	0.14	&	6.6	&	1.5	&	2500	&	653	&	1824	$^{+193}_{-{193}}$	&	3.8	&	2.39 	$^{+0.42}_{-{0.47}}$	&	1.98 	$^{+0.08}_{-{0.08}}$ &	\\
	\hline\noalign{\smallskip}
    \multicolumn{15}{l}{
    \footnotesize \textbf{Notes.} 
    \parbox[t]{0.9\textwidth}{\footnotesize
    $^1$ The lower threshold of the CO emission for on-cloud stars. $^2$ The number of the on-cloud stars used to calculate the distance. $^3$ The estimated distances of this work. $^4$ Total mass of molecular gas in molecular clouds estimated with the $^{12}$CO-to-H$_2$ mass conversion factor of X=2.0$\times$10$^{20}$ cm$^\mathrm{-2}$ (K km s$^{-1}$)$^{-1}$ \citep{Bauermeister+2013}, which only takes CO-bright components into account. $^5$ Kinematic distances derived from the A5 model of \citet{Reid+2014}. $^6$ Kinematic distances derived from the model in \citet{Reid+2019}.
    }
    } \\
\end{longtable}
\endgroup

In addition, two prominent star-forming regions, M16 and M17, are located within the observed PPV space. Taking M16 as an example, we detected 114 $^{13}$CO clouds that share both spatial and kinematic association with it. The distance to M16 has been well constrained in previous studies, with reported values including maser distance of 1495$^{+44}_{-{41}}$ pc \citep{Reid+2019}, extinction distance of 1699$^{+16}_{-{16}}$ pc \citep{Zucker+2020}, and open cluster (NGC 6611) distance of 1698$^{+5}_{-{5}}$ pc \citep{Hunt+2023}. We used these established estimates to assign distances to the associated $^{13}$CO clouds, as listed in Table \ref{Table_starformation}. Our catalog contains eight $^{12}$CO molecular clouds whose estimated distances are consistent with those of the star-forming regions M16 and M17. Among these, G017.5+02.3 is the largest cloud, with a length of about 30.1 pc and a mass of about $1.23 \times 10^{5}$ M$_{\odot}$, an average radial velocity of about 27.754 km s$^{-1}$, and a velocity dispersion of 2 km s$^{-1}$.

\begin{table*}
    \centering
    \renewcommand\arraystretch{1.3}  	
	\centering
	\setlength{\tabcolsep}{3.1mm}
    {   
    \caption{Distance of the $^{13}$CO clouds associated with the star-forming regions.}
    \label{Table_starformation}
    \begin{tabular}{cccccc}
    \hline
    \multirow{2}{*}{} & \multicolumn{4}{c}{$^{13}$CO clouds} & \multirow{2}{*}{D$_{\rm Lit}$ (pc)} \\
    \cmidrule(lr){2-5}
     & $l$ ($^{\circ}$) & $b$ ($^{\circ}$) & $V_{\rm LSR}$ (km\,s$^{-1}$) & Number & \\
    \midrule
    M16 & [15.829,18.109] & [0.037,1.414] & [18.0,27.9] & 114 & 1495$^{+44}_{-41}$ $^{\textcolor{blue}{a}}$,1699$^{+16}_{-16}$ $^{\textcolor{blue}{b}}$,1698$^{+5}_{-5}$ $^{\textcolor{blue}{c}}$ \\
    M17 & [15.029,15.364] & [$-$0.661,$-$0.404] & [16.1,23.7] & 15 & 2004$^{+99}_{-11}$ $^{\textcolor{blue}{a}}$,1516$^{+13}_{-13}$ $^{\textcolor{blue}{b}}$ \\
    \hline
    \multicolumn{6}{l}{
    \footnotesize \textbf{Notes.} 
    $^a$ \citet{Reid+2019} -- maser, $^b$ \citet{Zucker+2020} -- extinction method, $^c$ \citet{Hunt+2023} -- open cluster.} \\
    \end{tabular}}
\end{table*}

\begin{table*}
    \centering
    \renewcommand\arraystretch{1.2}  	
	\centering
	\setlength{\tabcolsep}{1.2mm}
    {   
    \caption{Comparison with literature distances.}
    \label{Table_ComparisonD}
    \begin{tabular}{>{\centering\arraybackslash}m{2cm}>{\centering\arraybackslash}m{2.cm} >{\centering\arraybackslash}m{3cm} >{\centering\arraybackslash}m{7cm}}
        \hline  
        ID & Name & D$_{\rm Our}$ & D$_{\rm Lit}$     \\
         &  &  (pc) & (pc)  \\
        (1) & (2) & (3) & (4) \\
        \hline
        1 & G$010.9+02.0$	&	1294	$^{+130}_{-{130}}$	&	1427$^{+34}_{-{34}}$	$^{\textcolor{blue}{b}}$	\\
        2 & G$011.2-02.2$	&	1302	$^{+130}_{-{130}}$	&	1250$^{+54}_{-{50}}$ $^{\textcolor{blue}{a}}$,1477$^{+35}_{-{35}}$ $^{\textcolor{blue}{b}}$,	1377$^{+16}_{-{21}}$  $^{\textcolor{blue}{c}}$,	1236$^{+75}_{-{72}}$ $^{\textcolor{blue}{d}}$	\\
        3 & G$013.8+03.9$	&	1821	$^{+184}_{-{184}}$	&	1841$^{+43}_{-{43}}$ $^{\textcolor{blue}{b}}$	\\
        4 & G$015.7-01.3$	&	1506	$^{+153}_{-{153}}$	&	1653$^{+39}_{-{39}}$	$^{\textcolor{blue}{b}}$	\\
        5 & G$016.5-03.1$	&	1895	$^{+192}_{-{192}}$	&	2268$^{+54}_{-{54}}$  $^{\textcolor{blue}{b}}$	\\
        6 & G$016.8-02.4$	&	2093	$^{+211}_{-{211}}$	&	2350$^{+647}_{-{417}}$	$^{\textcolor{blue}{a}}$, 2095$^{+17}_{-{26}}$	$^{\textcolor{blue}{c}}$	\\
        7 & G$017.5+02.3$	&	1409	$^{+141}_{-{141}}$	&	1507$^{+47}_{-{47}}$	$^{\textcolor{blue}{b}}$   \\
        8 & G$019.3-01.0$	&	1722	$^{+193}_{-{193}}$	&	1926$^{+45}_{-{45}}$	$^{\textcolor{blue}{b}}$	\\
        9 & G$019.4-03.7$	&	275	$^{+28}_{-{28}}$	&	270$^{+6}_{-{6}}$	$^{\textcolor{blue}{b}}$	\\
        \hline
        \multicolumn{4}{l}{
        \footnotesize \textbf{Notes.} 
        \parbox[t]{0.9\textwidth}{\footnotesize
        $^a$ \citet{Reid+2019} -- maser, $^b$ \citet{ChenBQ+2020a} and $^c$ \citet{Zucker+2020} -- extinction method, $^d$ \citet{Zhang+2023} -- young stellar objects (YSOs).
        }
        } \\
    \end{tabular}}
\end{table*}

\subsection{Comparison with previous distance results}
As listed in Table \ref{Table_ComparisonD}, we have collected distance estimates for nine $^{12}$CO clouds from the literature for comparison with our results. Overall, our results are in good agreement with the literature.

In this region, there are a total of 12 maser sources \citep{Reid+2019} with distances derived from trigonometric parallaxes located within the relevant l--b--V space. Two of these masers are associated with the G011.2$-$02.2 and G016.8$-$02.4 clouds, with distances of $1250^{+54}_{-50}$ pc and $2350^{+647}_{-417}$ pc, respectively. As demonstrated in Table \ref{Table_ComparisonD}, our estimated distances for clouds are $1302^{+130}_{-130}$ pc (G011.2$-$02.2) and $2093^{+211}_{-211}$ pc (G016.8$-$02.4), both consistent with the maser-based distances. Furthermore, other literature estimates for the distance to the G011.2$-$02.2 cloud are 1377$^{+16}_{-{21}}$ pc \citep{Zucker+2020}, 1477$^{+35}_{-{35}}$ pc \citep{ChenBQ+2020a}, and 1236$^{+75}_{-{72}}$ pc \citep{Zhang+2023}, all of which agree well with our current result. For the G016.8$-$02.4 cloud, the literature estimate of the distance for this component is 2095$^{+17}_{-{26}}$ pc \citep{Zucker+2020}, which is also consistent with our distance estimate. Additionally, distances for seven other molecular clouds reported by \citet{ChenBQ+2020a} are consistent with our results.

Except for the nine clouds listed in Table \ref{Table_ComparisonD}, we did not find accurate distance references for the remaining 47 molecular clouds.

\subsection{Comparison with kinematic distances}
\begin{figure*}
    \centering
    \includegraphics[width=1\linewidth]{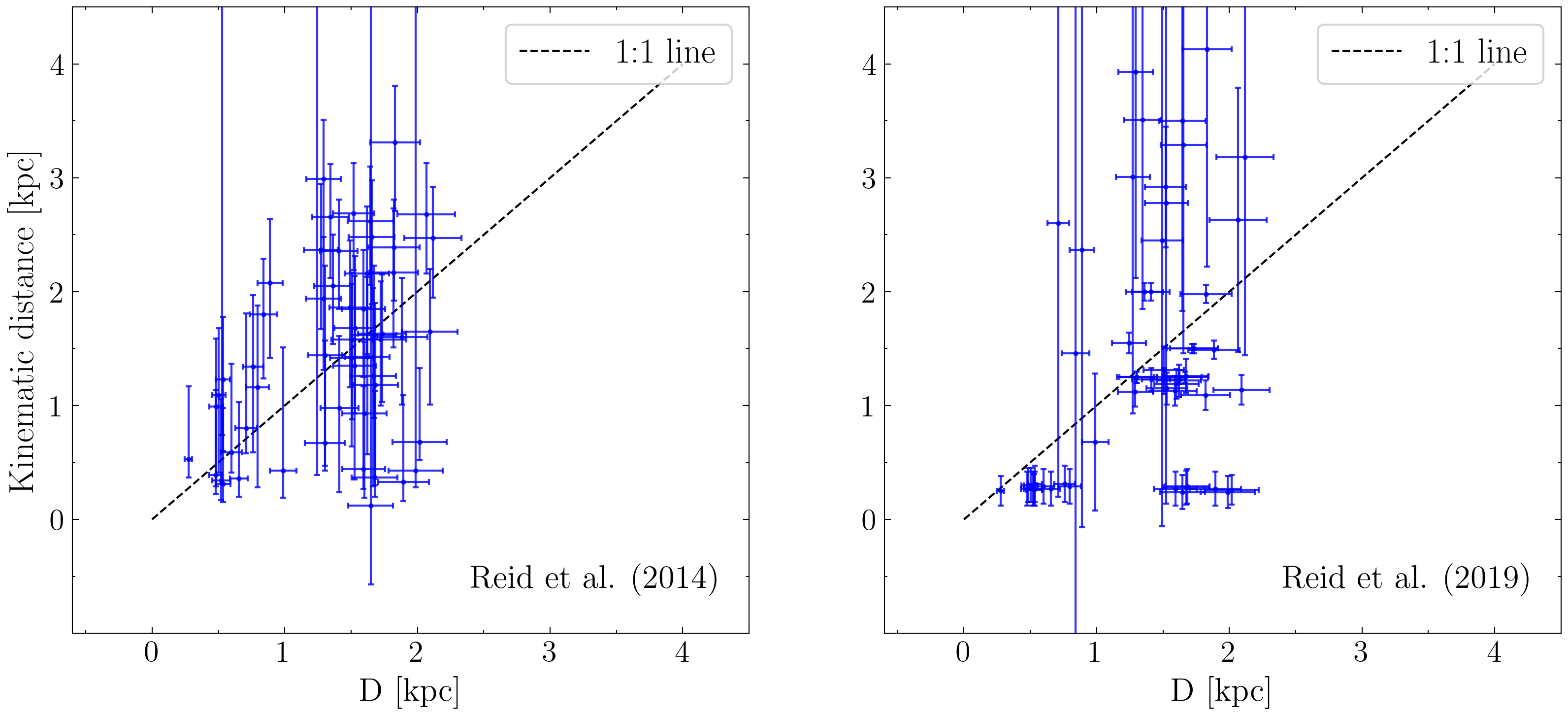}
    \caption{Comparison with kinematic distances derived from the A5 model in \cite{Reid+2014} and its improved version from \cite{Reid+2019}.}
    \label{Comparison_KD}
\end{figure*}

We compare our distance results with kinematic distances derived from the A5 model in \cite{Reid+2014}\footnote{\url{https://www3.mpifr-bonn.mpg.de/staff/abrunthaler/bessel_calc/revised_kd.html}} and its improved version from \cite{Reid+2019}\footnote{\url{https://www3.mpifr-bonn.mpg.de/staff/abrunthaler/bessel_calc2.0/}}, as displayed in Fig. \ref{Comparison_KD}. We adopted the near kinematic distance from both the \cite{Reid+2014} and \cite{Reid+2019} models for each cloud, by inputting its Galactic coordinates $(l, b)$ and V$_{\text{LSR}}$, while keeping all other model parameters at their default values. We note that for two clouds, G010.2$-$03.5 and G018.6$+$04.1, the recalculated near-kinematic distances still appear relatively large. Using the online kinematic distance calculator, we obtained near distances of $17.52^{+2.15}_{-17.13}$ kpc and $15.81^{+0.18}_{-15.07}$ kpc for these two clouds, and far distances of $17.52^{+2.15}_{-1.49}$ kpc and $15.22^{+0.76}_{-0.65}$ kpc, respectively. The extremely large and highly asymmetric lower uncertainties of the near distances reflect the well-known kinematic distance ambiguity inherent in the first Galactic quadrant. For these clouds, the near-distance solution is highly uncertain, and the possibility of a far-distance solution cannot be entirely ruled out based on kinematics alone.

As shown in the Fig. \ref{Comparison_KD}, kinematic distances are generally compatible with our distances within uncertainties for molecular clouds ($<$ 1 kpc). Kinematic distances have a major problem, as it is difficult to resolve the ambiguity between near and far distances. We found that a small fraction of clouds at distances $>1~\mathrm{kpc}$ have estimated distances that are significantly smaller than their kinematic distances, suggesting that the motions of molecular clouds beyond $1,\mathrm{kpc}$ may deviate from the Galactic rotation curve. 
Nevertheless, most kinematic distances are compatible with our distances within the uncertainties.

\subsection{The physical properties and spatial distribution of molecular clouds}
\begin{figure*}
    \centering
	\begin{subfigure}{0.51\linewidth} 
        \centering
        \includegraphics[width=\linewidth]{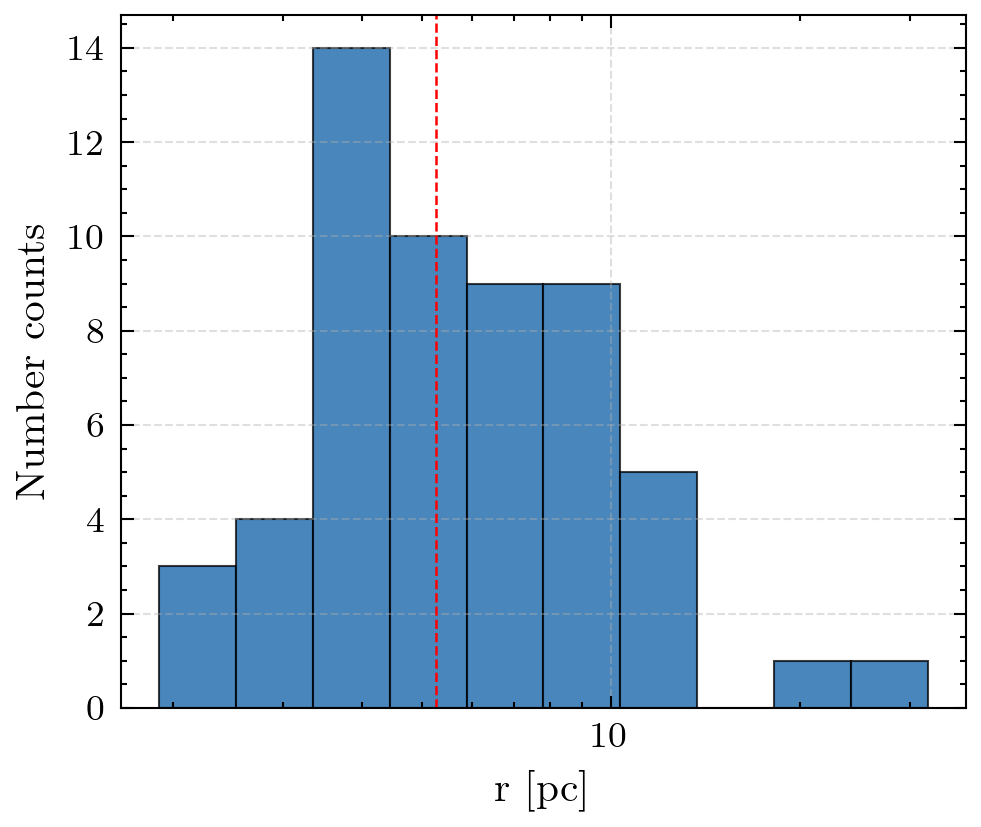}
        \label{Hist_rcpc}
    \end{subfigure}
    \hfill 
    \begin{subfigure}{0.48\linewidth} 
        \centering
        \includegraphics[width=\linewidth]{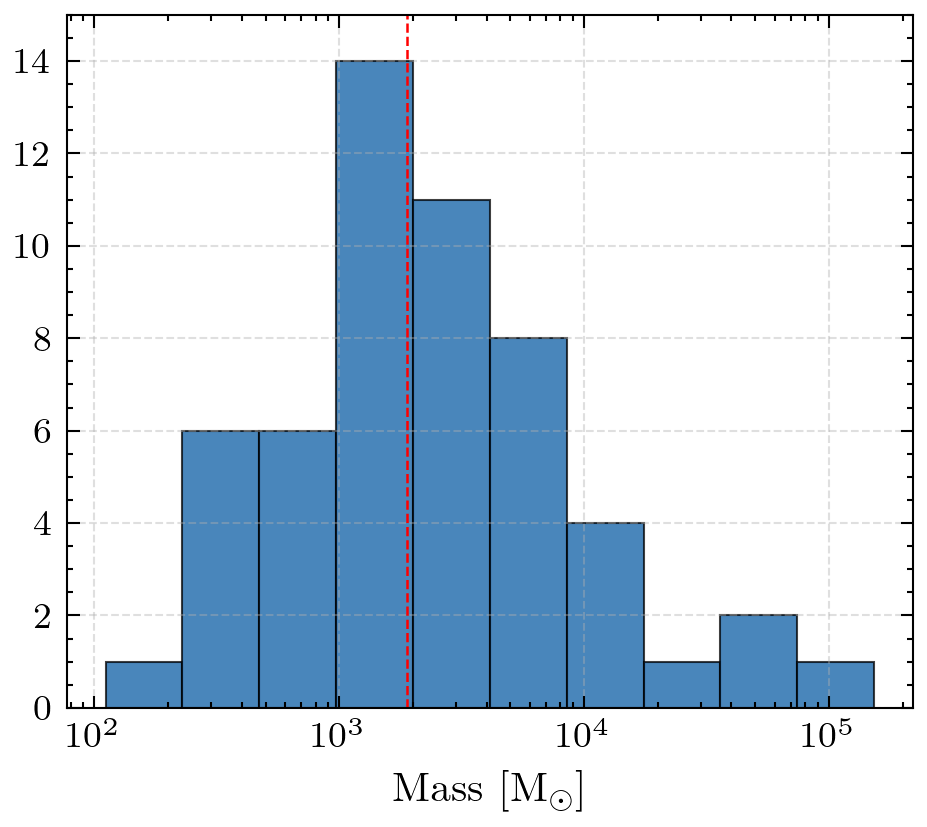}
        \label{Hist_mass}
    \end{subfigure}
    \caption{The histogram distributions of the physical properties, including linear radius (${\it Left}$) and masses (${\it Right}$) of our cataloged molecular clouds. The red vertical dashed lines are median values.}
    \label{Hist_rcpc_mass}
\end{figure*}

\begin{figure*}
    \centering
    \includegraphics[width=0.8\linewidth]{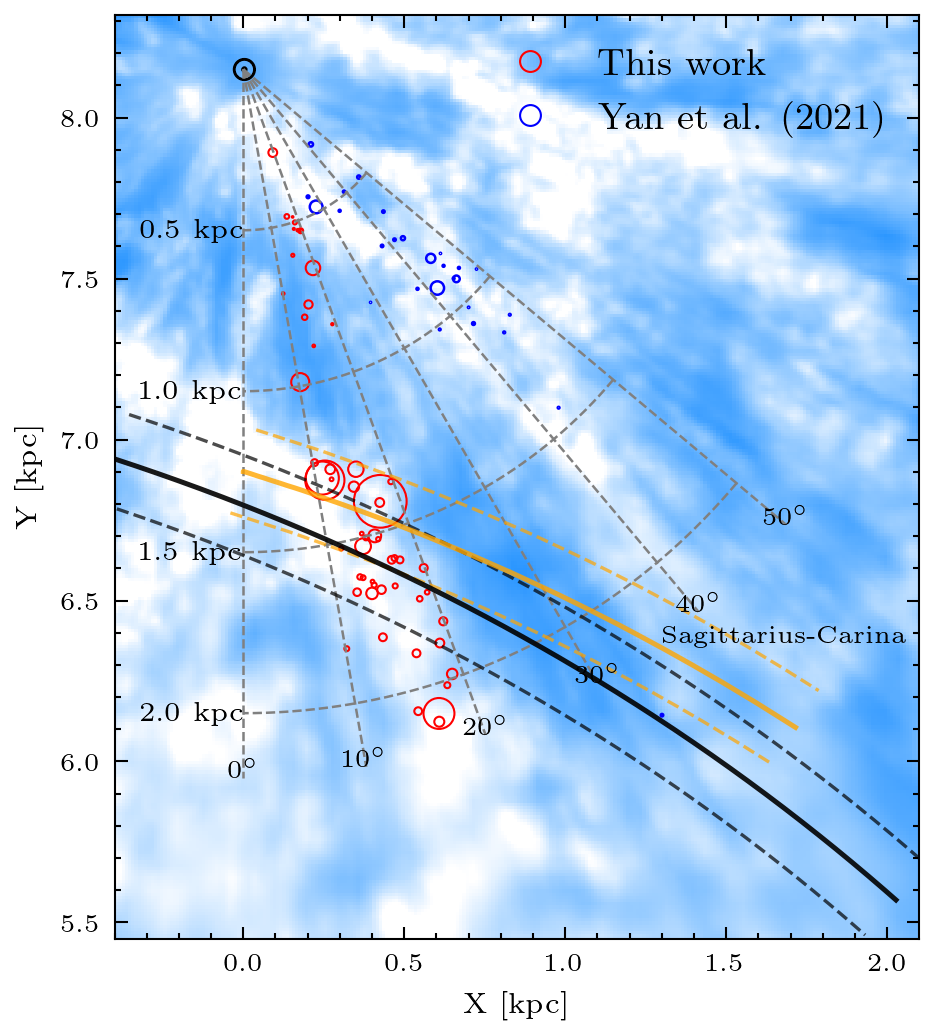}
    \caption{Spatial distribution of the molecular clouds identified in the current work (red hollow circles) and \citet[blue]{Yan+2021} in the Galactic coordinates. The background is an extinction map \citep{Vergely+2022} of the Galactic dust of disc vertical height $|Z| < 0.4 \, \text{kpc}$. The origin of the coordinate is the Galactic center, and the marker sizes are scaled with the mass. The distance of the Sun to the Galactic center is 8.15 kpc. The orange lines represent Sagittarius-Carina Arm from \cite{Reid+2019}. The black lines indicate the Sagittarius-Carina Arm from \cite{Xu+2023}.}
    \label{spatial-distribution}
\end{figure*}

The derived physical properties of the individual clouds, linear radius $r$ and masses, are also listed in Table \ref{Catalog}. The distributions of the linear radius and masses of the molecular clouds are presented in Fig. \ref{Hist_rcpc_mass}, and the red dashed lines are median values. For the cataloged clouds, their linear radius ranges from $\sim$ 1.9 to 30.1 pc, with a median value of 5.25 pc, and their masses range from $\sim$ 1.54$\times$10$^2$ to 1.6$\times$10$^5$ M$_{\odot}$, with a median value of 1.9$\times$10$^3$ M$_{\odot}$.
\cite{Vergely+2022} presented a 3D distribution of the interstellar dust within a volume of 6$\times$6$\times$0.8 kpc$^3$ the Sun, derived using a hierarchical inversion algorithm based on the photometric data of \textit{Gaia} DR2 and 2MASS, and the parallaxes of \textit{Gaia} EDR3. As shown in Fig. \ref{spatial-distribution}, we plot the spatial distribution of our molecular clouds together with the 3D dust distribution of the Galactic plane from \cite{Vergely+2022}. The distribution of 27 molecular clouds from \cite{Yan+2021} is also overplotted in the Figure \ref{spatial-distribution}. The spatial distribution of the molecular clouds cataloged here is in good agreement with the dust distribution of \cite{Vergely+2022}. The consistency supports the robustness of the methods used in both studies.
By combining the spatial, mass, and size distributions of the molecular clouds, we find that most of the molecular clouds with distance estimates are relatively small-scale and distributed between the spiral arms, while larger molecular clouds are generally located along the spiral arms. Compared with dust extinction maps, our method allows us to detect more molecular clouds and to estimate their distances.

\begin{figure*}
    \centering
    \includegraphics[width=0.65\linewidth]{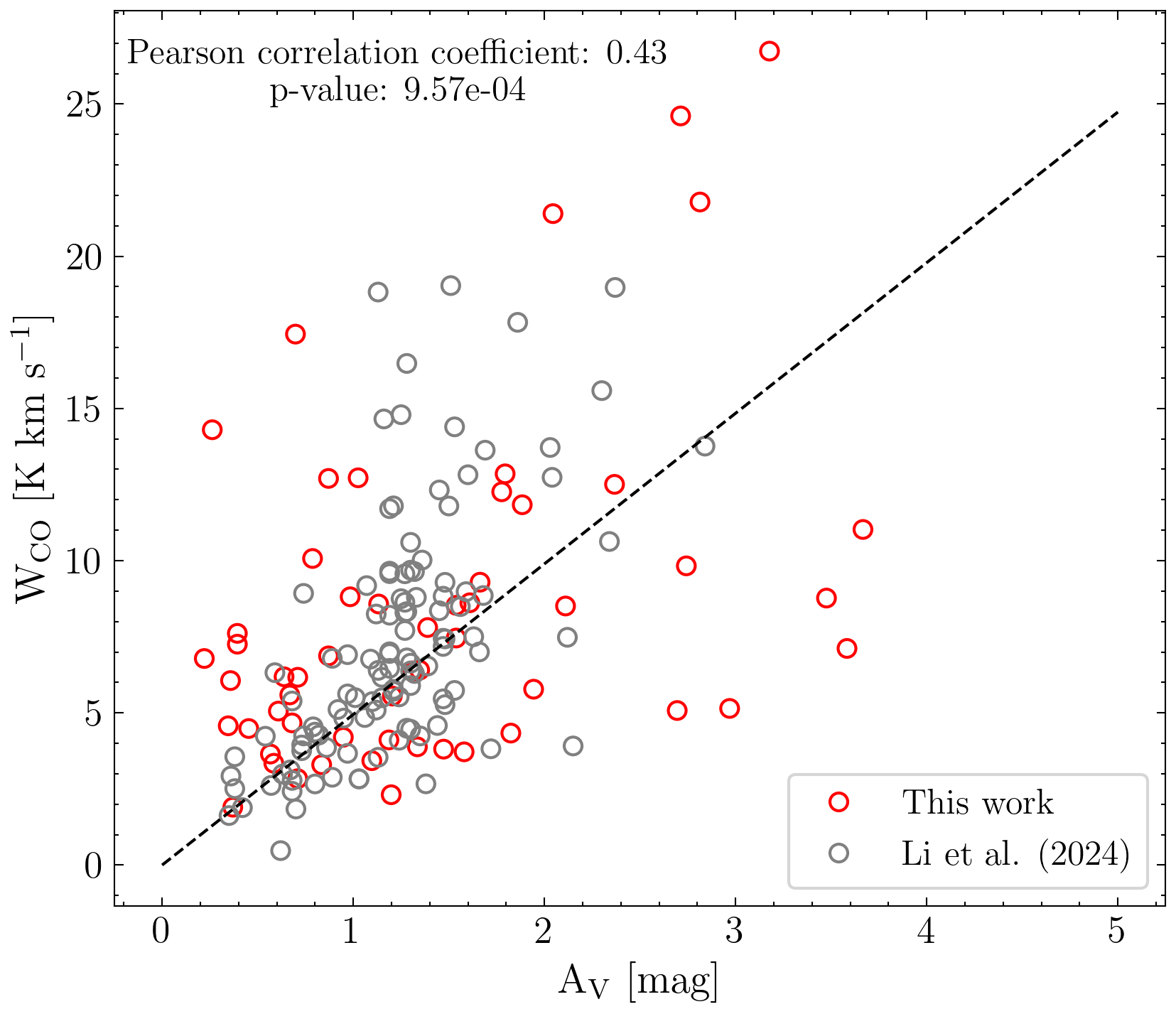}
    \caption{The correlation between the A$_\mathrm{V}$ and W$_{\mathrm{CO}}$ of our cataloged molecular clouds (red hollow circle), and of those from \citet[gray]{Li+2024}. The black dashed line shows the relationship implied by the conversion between W$_{\mathrm{CO}}$ and H$_2$ derived from $\gamma$-rays \cite{Strong+1996}.
    }
    \label{AVvsW12CO_fit}
\end{figure*}

After determining the molecular cloud distance, the average color excess $\overline{E(J-K_s)}$ can be obtained from the mean ($\mu_1$) of the foreground star colors and the mean ($\mu_2$) of the background star colors, defined as $\overline{E(J-K_s)}$=$\mu_1$-$\mu_2$. This color excess is converted to visual extinction A$_\mathrm{V}$ using the relation $\frac{E(J-K_s)}{E(B-V)}$ $\approx$ 0.52 \citep{Wang+2019}. We assume a total-to-selective extinction ratio R$_{\rm V}=3.1$, which yields A$_\mathrm{V}$ $= R_{\rm V} \cdot \overline{E(J-K_s)}/0.52$. For the velocity-integrated $J=1-0$ emission (W$_{\mathrm{CO}}$), we take the average of the intensity that exceeds an intensity threshold (CO$_{\rm cut}$) for each cloud. The relationship between CO integrated intensity (W$_{\mathrm{CO}}$) and visual dust extinction (A$_\mathrm{V}$) is shown in Fig. \ref{AVvsW12CO_fit}. As a convenient reference, the black dashed line corresponding to the conversion of W$_{\mathrm{CO}}$ to A$_\mathrm{V}$, N$_\mathrm{H_2}$ = 2 $\times$ 10$^{20}$ W$_{\mathrm{CO}}$ cm$^{-2}$ (K km s$^{-1}$)$^{-1}$ and N$_\mathrm{H_2}$ = 9.4 $\times$ 10$^{20}$ A$_\mathrm{V}$ cm$^{-2}$ is also drawn on the plots. The result of the Pearson correlation test is shown in the plot, with a correlation coefficient of 0.43 and a p-value of $9.57 \times 10^{-4}$. It is evident that there is a moderate correlation between W$_{\mathrm{CO}}$ and A$_\mathrm{V}$. Recently, \cite{Li+2024} found a good linear relationship between the gas and dust in 112 strongly correlated clouds, where A$_\mathrm{V}$ is derived from the average color excess within the dust cloud. We have plotted their results in Fig. \ref{AVvsW12CO_fit}, which are consistent with our results, and both show an upward trend in $W_{\mathrm{CO}}$ around $A_\mathrm{V} \approx 1.2$.

\section{Summary}  
We measured distances to molecular clouds within the Galactic longitude range $l=10\degr-20\degr$ and latitude $\lvert b \rvert \leq 5.25\degr$. We identified 216 $^{12}$CO molecular clouds using the DBSCAN algorithm and directly estimated distances for 56 of them, with distances ranging from $\sim$275 to $\sim$2118 pc. Among these, 47 are accurately determined for the first time. The typical statistical uncertainty of the distance is $\sim$ 5$\%$, and the systematic uncertainty is $\sim$ 10$\%$. We have also measured the physical properties of the clouds, such as linear radius and mass. The spatial distribution of these clouds is consistent with that of the dust extinction distribution. We also found a moderate correlation between the dust extinction and the $^{12}$CO integrated intensity.

\begin{acknowledgements}
This work is supported by the National Key R\&D Program of China (grant No. 2023YFA1608000). ZC acknowledges the Natural Science Foundation of Jiangsu Province (grants No. BK20231509). This research made use of the data from the Milky Way Imaging Scroll Painting (MWISP) project, which is a multi-line survey in $^{12}$CO/$^{13}$CO/C$^{18}$O along the northern galactic plane with PMO-13.7m telescope. We are grateful to all the members of the MWISP working group, particularly the staff members at the PMO-13.7 m telescope, for their long-term support. MWISP is sponsored by the National Key R\&D Program of China with grants 2023YFA1608000 and 2017YFA0402701, and the CAS Key Research Program of Frontier Sciences with grant QYZDJ-SSW-SLH047. This work has made use of data from the European Space Agency (ESA) mission Gaia (\url{https://www.cosmos.esa.int/gaia}), processed by the Gaia Data Processing and Analysis Consortium (DPAC, \url{https://www.cosmos.esa.int/web/gaia/dpac/consortium}). Funding for the DPAC has been provided by national institutions, in particular the institutions participating in the Gaia Multilateral Agreement. This publication makes use of data products from the Two Micron All Sky Survey, which is a joint project of the University of Massachusetts and the Infrared Processing and Analysis Center/California Institute of Technology, funded by the National Aeronautics and Space Administration and the National Science Foundation.
\end{acknowledgements}

\bibliographystyle{raa}
\bibliography{RAA-2025-0581}

\newpage
\section*{Appendix}  
\section{The color-distance diagram for cloud-free areas}

\end{CJK*}
\end{document}